\documentclass[a4paper,12pt]{article}

\usepackage{jheppub} 
\usepackage{lineno}
\usepackage{amsmath}
\usepackage{mathtools}

\makeatletter
\gdef\@fpheader{}
\makeatother

\title{\boldmath Numerical Computations of Entanglement Measures in Curved Space}

\author[a]{Suresh Govindarajan}
\author[a]{Sreehari A Padinhareveettil}
\author[b]{Raghotham A Kulkarni}
\affiliation[a]{Department of Physics, Indian Institute of Technology Madras \\ Chennai 600036 India}
\affiliation[b]{Centre for High Energy Physics,
Indian Institute of Science,
C. V. Raman Avenue,\\ Bangalore 560012, India.}
\emailAdd{suresh@physics.iitm.ac.in\\ apsreehari97@physics.iitm.ac.in\\ raghothamk@iisc.ac.in}

\abstract{We numerically compute the entanglement entropy and negativity for scalar fields and abelian gauge fields in a variety of situations. These extend computations of Srednicki to situations involving curved space. We discretize space in a covariant way. Finally, we compare some of our results with those obtained via the heat kernel coefficients.}

\begin{document}
\maketitle
\flushbottom
\pagebreak

\pagenumbering{arabic}
\section{Introduction}
The works of Bombelli et al. and Srednicki \cite{PhysRevD.34.373,Srednicki:1993im} studied the entanglement entropy of quantum fields. While the focus of Bombelli et al. was on the contribution to the black hole entropy, Srednicki performed the calculation for fields in flat space and showed that the entanglement entropy followed the area law, i.e. it is proportional to the area of the entangling surface. In Srednicki's work, this was taken to be a sphere and the state considered was the ground state of the discretized field inside the sphere. Srednicki also noted the importance of his result in the context of black hole entropy and these results found great importance in the subsequent works of Page on black hole evaporation \cite{Page:1993up,Page:1993df,Page:1993wv}. 

The analytical method that we have to compute the entanglement entropy is the replica trick. While this is supposed to be an easy alternative to computing the log of the density matrix, constructing the replica manifold is difficult and the calculation can be done exactly only in some simple cases, such as the one for two-dimensional conformal field theories which was performed by Cardy and Calabrese \cite{Calabrese:2012ew}. However, the numerical method can be adapted and used in various backgrounds and there are many works that have explored such directions \cite{Casini:2015dsg,Lohmayer:2009sq,Benedetti:2019uej,Boutivas:2024lts,Boutivas:2025rdf}. This is what the bulk of this work also deals with. Specifically, we use the method to study entanglement entropy in AdS spacetime for spherical entangling surfaces. We also  do the same for fields inside the RT surface \cite{Ryu:2006bv,Ryu:2006ef} in AdS$_3$. The entanglement entropy of bulk fields in AdS is of great interest in AdS/CFT because it contributes to the total entanglement entropy of the subsystem of the dual CFT. This is explored in various proposals such as the FLM formula, quantum extremal surfaces and islands \cite{Faulkner:2013ana,Engelhardt:2014gca,Almheiri:2019hni,Almheiri:2019yqk}. We further extend the numerical method to calculate the logarithmic negativity, once again, working with the ground state. This is made possible by the fact that the negativity takes a simple form in terms of the eigenvalues of the reduced density matrix when the state is pure. Specifically, this is simply given by the Renyi entropy of order $1/2$. While we focus on numerical calculations for the bulk of this work, we look at one analytic tool, namely the heat kernel method \cite{Fursaev:1995ef,Solodukhin:2011gn,Nishioka:2018khk}, which gives the entanglement entropy as a series of divergent terms containing the UV cut-off of the theory. Out of these divergent terms, there is one universal term which is logarithmically divergent in the UV cut-off in even spacetime dimensions and constant in odd spacetime dimensions. In the case of even dimensions, the heat kernel method allows us to analytically calculate the coefficient of the universal term. While we know that the leading order divergent term obeys the area law, its coefficient is scheme-dependent and does not mean anything as such. 

In this work, we calculate the entanglement entropy for scalar and vector fields in global AdS$_4$ and within RT surfaces in AdS$_3$ using the numerical method of Srednicki. In each of the above cases, we find the suitable basis for the fields such that we can integrate the coordinates from the Hamiltonian except the one which we will discretize. From the results of the calculation, we confirm the expected area law behaviour. For scalar field in AdS$_4$, we see that the entanglement entropy has a dependence of the AdS radius $L$. We also apply the same technique to compute logarithmic negativity in 2+1 dimensions and see that the result is proportional to the area. This calculation cannot be done in higher dimensions because of a divergence arising from the angular coordinates. This is also true for entanglement entropy in higher dimensions \cite{Srednicki:1993im}.
We also compute the coefficient of the universal term in the entanglement entropy of scalar field in global $AdS_4$ and confirm that the known flat space value is recovered in the $L\rightarrow\infty$ limit.

The paper is organized as follows. In section 2, we look at some coordinate systems for AdS, which we will use later in the calculations. In Section 3, we summarize Srednicki's method of numerical calculation and extend the calculations in the presence of mass. In section 4, we calculate the entanglement entropy for both scalar and vector fields in a hyperspherical surface in AdS$_4$. For the calculation for vector fields, we use a version of vector spherical harmonics suitable for AdS$_4$. In section 5, we study the entanglement entropy of scalar and vector fields inside an RT surface in AdS$_3$. In section 6, we use the same method to calculate the entropy for a scalar fields in dS$^4$. In section 7, we show how Srednicki's method can be used to calculate logarithmic negativity and demonstrate it for fields in 2+1 dimensions, in flat space and in AdS$_3$ for fields inside the RT surface. In section 8, we review the heat kernel method and use it to calculate the leading order behaviour of the enganglement entropy and reconfirm our numerical results. We also calculate the coefficient of the universal term in the case of even spacetime dimensions. \\
During the completion of this work, the authors became aware of a work along similar lines  \cite{Boutivas:2025ksp}. One important difference in method that we have used is covariant spacing, or using the proper distance as the variable to be discretized. We have also tried to apply the calculations to vector fields in different backgrounds. Moreover, while we have not pushed the numerical precision so as to obtain the coefficients of subleading terms, we have done these calculations explicitly using the heat kernel method.
\section{Coordinates for AdS$_4$}
The four dimensional Anti de Sitter spacetime can be embedded in a $\mathbb{R}^{2,3}$:
\begin{equation}
    -(T_1)^2-(T_2)^2+(X_1)^2+(X_2)^2+(X_3)^2=-L^2
\end{equation}
The metric is:
\begin{equation}
    ds^2=-dT_1^2-dT_2^2+dX_1^2+dX_2^2+dX_3^2
\end{equation}
\subsection{Global Coordinates}
The global coordinates for AdS are defined as follows:
\begin{align}
    T_1=&\sqrt{r^2+L^2} \, \text{cos}\Big(\frac{t}{L}\Big) \\
    T_2=&\sqrt{r^2+L^2} \, \text{sin}\Big(\frac{t}{L}\Big) \\
    X_3=& \ r \, \text{cos} \phi\\
    X_1=& \ r \, \text{sin} \phi \, \text{cos} \theta\\
    X_2=& \ r \, \text{sin} \phi \, \text{sin} \theta
\end{align}
Here, the radial coordinate takes the values $r\in[0,\infty)$, and the hyperspherical coordinates on $S^{d-1}$ take the values $\phi\in[0,2\pi)$, $\theta \in[0,\pi]$. We see that there are closed timelike curves in the coordinates defined above. We avoid this by  passing into the universal covering space. In these coordinates, the metric takes the form:
\begin{equation}\label{globmetric}
   ds^2=-dt^2\Big(1+\frac{r^2}{L^2}\Big)+\frac{dr^2}{1+\frac{r^2}{L^2}}+r^2 d\Omega_2^2 
\end{equation}
The boundary of the spacetime is at $r=\infty$, where $ds^2\approx -r^2 \, dt^2+r^2 \, d\Omega_{2}^2$, which corresponds to the geometry of $S^{2}\times\mathbb{R}$.

\subsection{Foliation using geodesics}
\begin{figure}[h]
    \centering
    \includegraphics[scale=3.75]{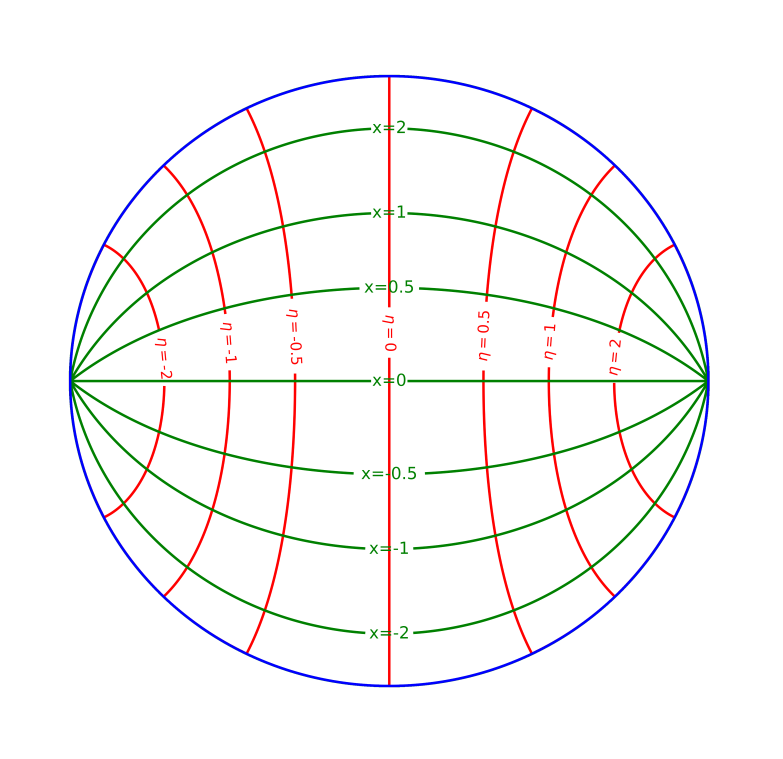}
    \caption{Foliation of a constant $t$ slice using constant $\eta$ and $x$ surfaces. $L$ is taken to be 1 here. Constant $\eta$ surfaces (red) are RT surfaces. Both $\eta$ and $x$ can take values from $-\infty$ to $\infty$.}
    \label{x eta}
\end{figure}

Another set of coordinates that are particularly useful for the calculation of the entanglement entropy of fields inside RT surfaces are:
\begin{align}
    \eta \ = \ & L\, \text{sinh}^{-1}\Bigg(\frac{r \text{cos}\phi}{\sqrt{1+r^2 \text{sin}^2}\phi}\Bigg)\\
    x \ = \ & L\, \text{sinh}^{-1}(r \, \text{sin}\phi)
\end{align}
Both $\eta$ and $x$ can take values $(-\infty,\infty)$. We keep the $t$ in global coordinates unmodified. The metric then becomes
\begin{equation}\label{RTmetric}
    ds^2=- \ \text{cosh}^2 (x/L) \text{cosh}^2(\eta/L) dt^2 +  \text{cosh}^2(x/L) d\eta^2 + dx^2
\end{equation}
Finally, we use dimensionless coordinates $\tilde{t}=\frac{t}{L}$, $\tilde{\eta}=\frac{\eta}{L}$ and $\tilde{x}=\frac{x}{L}$.
\begin{equation}\label{rtfoliation}
    ds^2=- \ L^2\cosh^2 \tilde{x} \,\cosh^2\tilde{\eta}\, d\tilde{t}^2 +  L^2\cosh^2 \tilde{x}\, d\tilde{\eta}^2 + L^2 d\tilde{x}^2
\end{equation}
This metric can also be derived from the following definitions in terms of the coordinates of the four-dimensional embedding space:
\begin{equation}
\begin{split}
        X_0=& L\, \cosh\tilde{x} \, \cosh \tilde{\eta} \, \sin \tilde{t} \\
        X_1=& L\, \cosh \tilde{x} \, \cosh \tilde{\eta} \, \cos \tilde{t} \\
        X_2=& L\, \cosh \tilde{x} \, \sinh \tilde{\eta} \\
        X_3=& L\, \sinh \tilde{x} 
\end{split}
\end{equation}
Then, the metric can be derived from:
\begin{equation}
    ds^2=-dX_0^2-dX_1^2+dX_2^2+dX_3^2.
\end{equation}
Constant $\eta$ surfaces are geodesics (RT surfaces) (Figure \ref{x eta}). These coordinates are closely related to AdS-Rindler coordinares. In the AdS-Rindler coordinates, $\tilde{\eta}$ is treated as a parameter, not a coordinate) that determines the size of the Rindler patch and takes values in the range $[0,\infty)$. It is related to the size of an RT surface of angular size $\theta$ by
\begin{equation}
	\cosh\tilde{\eta}=\frac{1}{\sin\big(\frac{\theta}{2}\big)}
\end{equation}
AdS-Rindler coordinates and their relation with the coordinates above are explained in detail in appendix \ref{appA}.

\section{Area Law of Entanglement Entropy}
\subsection{Entanglement Entropy of Scalar Fields in 3+1 d flat spacetime}\label{srednicki}
\begin{figure}[h]
    \centering
    \includegraphics[scale=0.55]{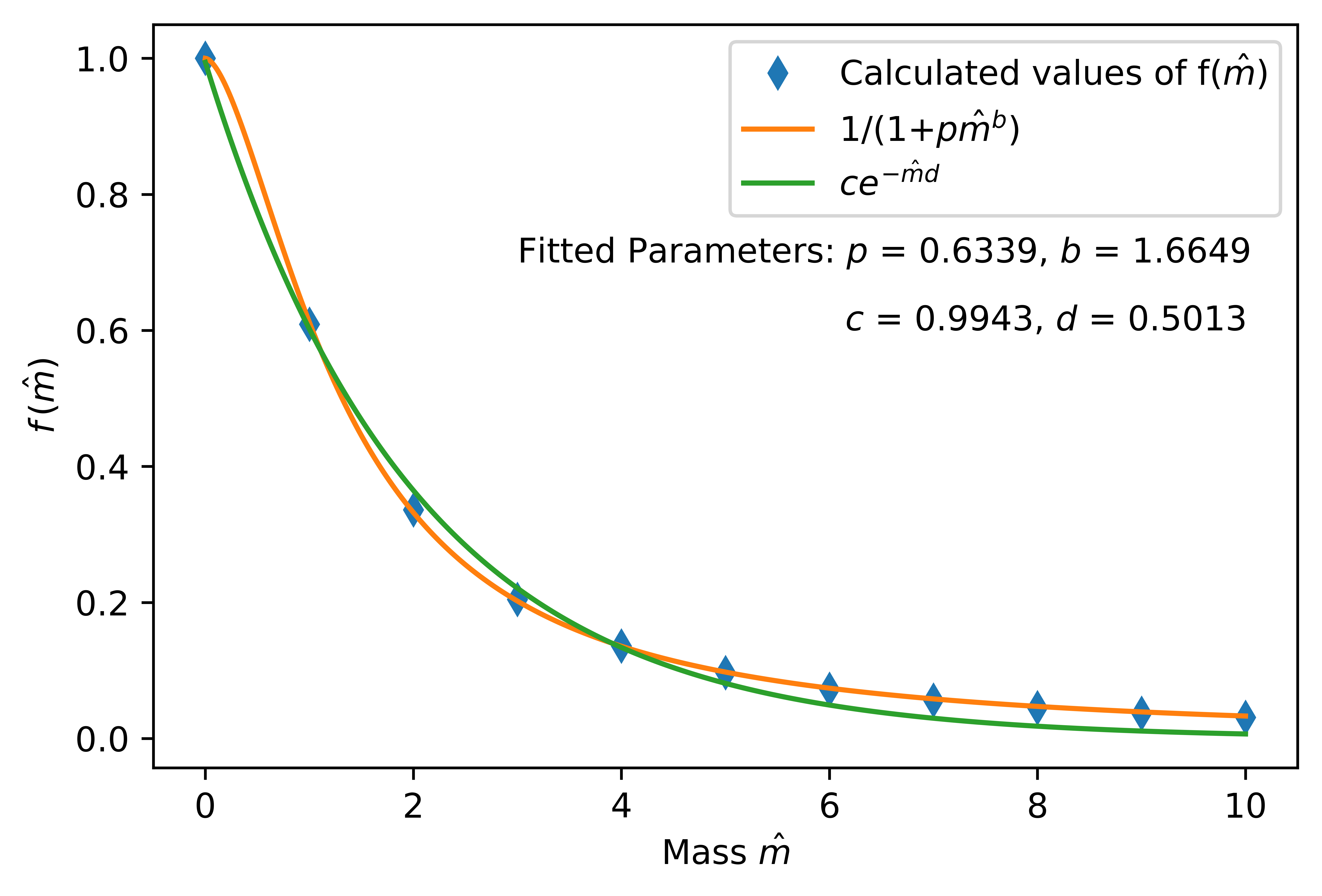}
    \caption{Fitting the dependence of entanglement entropy on mass (i) $1/(1+p\hat{m}^b)$ with $p=0.6339, \ b=1.6649$, (ii)$ce^{-\hat{m}d}$ with $c=0.9943,\ d= 0.5013$.}
    \label{flatmassfit}
\end{figure}

Srednicki showed through a numerical computation that the entanglement entropy of scalar fields inside a spherical surface in flat space is proportional to the area of the surface\citep{Srednicki:1993im}. Consider two coupled harmonic oscillators with the Hamiltonian:
\begin{equation}\label{csho}
    H=\frac{1}{2}\Big[p_1^2+p_2^2+k_0(x_1^2+x_2^2)+k_1(x_1-x_2)^2\Big]
\end{equation}
Changing to a new basis, $x_\pm=\frac{x_1+x_2}{\sqrt{2}}$, the Hamiltonian can be written as:
\begin{equation}
    H=\frac{1}{2}\Big[p_+^2+p_-^2+k_0 x_+^2+ (k_0+2k_1)x_-^2\Big]
\end{equation}
The ground state wavefunction is
\begin{equation}\label{gswavefn}
    \psi_0(x_1,x_2)=\pi^{-1/2}(\omega_+ \omega_-)^{1/4}\text{exp}[-(\omega_+ x_+^2+\omega_- x_-^2)],
\end{equation}
where $\omega_+=\sqrt{k_0}$ and $\omega_-=\sqrt{k_0+2k_1}$. Now we can write down the density matrix and integrate out one of the degrees of freedom to get the reduced density matrix.
\begin{equation}\label{gsrdm}
\begin{aligned}
    \rho_{\text{out}} (x_2,x_2')&=\int dx_1 \psi_0(x_1,x_2)\psi_0^*(x_1,x_2')\\
    &=\pi^{-1/2}(\gamma-\beta)^{1/2} \text{exp}[-\gamma(x_2^2+x_2'^2)/2+\beta x_2 x_2']
\end{aligned}
\end{equation}
where $\beta=\frac{(\omega_+-\omega_-)^2}{4(\omega_+ + \omega_-)}$ and $\gamma-\beta=2\frac{\omega_+ \omega_-}{\omega_+ + \omega_-}$. Eigenvalues and eigenfunctions of the reduced density matrix are:
\begin{align}\label{evals}
    p_n=&(1-\xi)\xi^n\\
    f_n(x)=&H_n (\alpha^{1/2} x) \text{exp} (-\alpha x^2/2)
\end{align}
such that 
\begin{equation}
    \int_{\infty}^{\infty}dx' \rho_{\text{out}}(x,x')f_n(x')=p_n f_n(x).
\end{equation}
$H_n(\alpha^{1/2} x)$ above denotes Hermite polynomials, $\alpha=(\gamma^2-\beta^2)^{1/2}$ and $\xi=\beta/(\gamma+\alpha$).
Now the von Neumann entropy can be calculated.
\begin{equation}
\begin{aligned}
    S&=\sum_n p_n \, \text{log} \, p_n\\
    &=-\text{log}(1-\xi)-\frac{\xi}{1-\xi}\text{log} \, \xi
\end{aligned}  
\end{equation}
This procedure can be generalized for a system of $N$ oscillators with the Hamiltonian of the form
\begin{equation}\label{HinK}
    H=\frac{1}{2}\sum_{i=1}^N p_i^2 + \frac{1}{2}\sum_{i,j=1}^N x_i K_{ij} x_j
\end{equation}
The ground state is given by:
\begin{equation}
    \psi_0(x_1,...,x_N) = \pi^{-N/4}(\text{det} \Omega)^{1/4}\text{exp}(-x^T\cdot\Omega\cdot x/2),
\end{equation}
where $\Omega=K^{1/2}$. Writing $\Omega$ as:
\begin{equation}
    \Omega = \begin{pmatrix}
        A & B \\
        B^T & C
    \end{pmatrix}
\end{equation}
The quantities analogous to $\beta$ and $\gamma$ in the two oscillator case can now be written down as matrices in terms of the submatrices of $\Omega$. 
We now consider scalar field in flat spacetime in 3+1 dimensions, inside a sphere of radius $r$. The solution for the equations of motion is of the form $\phi_{\ell m}(r)Y_{\ell}^m(\theta,\phi)$. We integrate out the spherical coordinates. The field and conjugate momenta are redefined so as to write down the Hamiltonian in a form similar to \eqref{HinK}.
\begin{equation} 
    H=\sum_{\ell m}\frac{1}{2} \int_0^{\infty} dr \bigg\{ \pi_{\ell m}^2(r) + r^2 \bigg[\frac{d}{dr}\bigg(\frac{\phi_{\ell m}(r)}{r}\bigg)\bigg]^2 +\frac{\ell(\ell+1)}{r^2}\phi_{\ell m}^2(r) \bigg\}
\end{equation}
We discretize the radial coordinate with spacing $a$ so that $r_{\text{max}}=Na$. Set $\phi_{\ell m,j}=\phi_{\ell m}(ja)$ and $\pi_{\ell m,j}=a\,\pi_{\ell m}(ja)$ . 
\begin{equation}\label{flathamiltonian}
    H_{\ell m}=\frac{1}{2a} \sum_{j=1}^{N} \bigg[ \pi_{\ell m,j}^2 + \bigg(j+\frac{1}{2}\bigg)^2 \bigg(\frac{\phi_{\ell m,j}}{j}-\frac{\phi_{\ell m,j+1}}{j+1}\bigg)^2 +\frac{\ell (\ell+1)}{j^2}\phi_{\ell m,j}^2 \bigg]
\end{equation}
In all the calculations that follow, the value of $a$ is set to 1 unless specified otherwise. From equation \eqref{flathamiltonian}, we see that the problem has now effectively become that of $N$ coupled harmonic oscillators. By integrating over the degrees of freedom from $r=0$ to $R=na$ where $n<N$, we can find the entanglement entropy of fields within a sphere of radius $R$.
Summing over $m$ gives a multiplicity of $2 \ell+1$ for each eigenvalue of the reduced density matrix. At large values of $\ell$, the eigenvalue $\xi_{\ell}$ is of the form: 
\begin{equation}\label{bigleval}
\xi_{\ell}(n)=\frac{n(n+1)(2n+1)^2}{64\ \ell^2(\ell+1)^2}+O(\ell^{-6}),
\end{equation}
and the entanglement entropy is:
\begin{equation}\label{biglentropt}
S(n,\ell)\simeq \xi_{\ell}(n)\Big[-\text{log}\xi_{\ell}(n)+1\Big].
\end{equation}
This shows that the infinite sum over $\ell$ will converge. It is seen numerically that the entanglement entropy depends on $R^2$, which means that the entropy grows with area of the spherical surface. In the two coupled harmonic oscillator case, it is seen that the entanglement entropy is a function of $k_1/k_0$. Similarly, for large values of the index in this case, the qualitative behaviour of the entanglement entropy can be found from the Hamiltonian, which will be proportional to the ratio of the coefficients of the derivative term and the quadratic terms of the field. Thus, we can read off the behaviour of the entanglement to be proportional to $R^2$.The slope obtained from the numerical calculation for the entropy vs. $R^2$ is $\kappa\sim 0.294$.
\begin{figure}[h]
    \centering
    \includegraphics[scale=0.6]{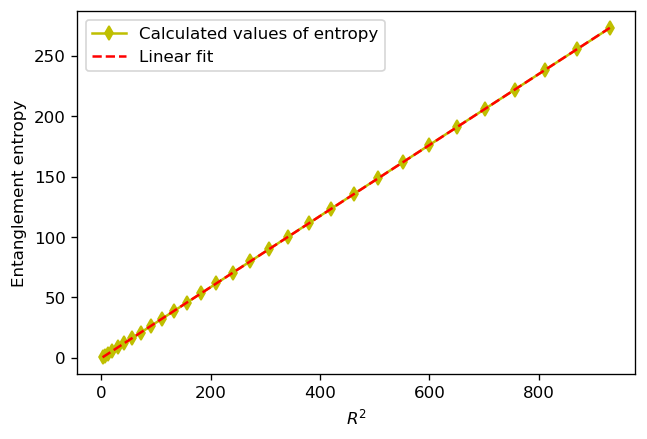}
    \caption{Entanglement entropy of scalar field in a sphere in flat space time vs. radius. Entanglement entropy increases linearly with $R^2$.}
    \label{flatspacemass}
\end{figure}
The procedure can be extended to massive scalar fields by modifying the third term in \eqref{flathamiltonian} to $\Big(\frac{\ell(\ell+1)}{j^2}+\hat{m}^2\Big)\phi_{\ell m,j}^2$. 
\begin{figure}
    \centering
    \includegraphics[scale=0.6]{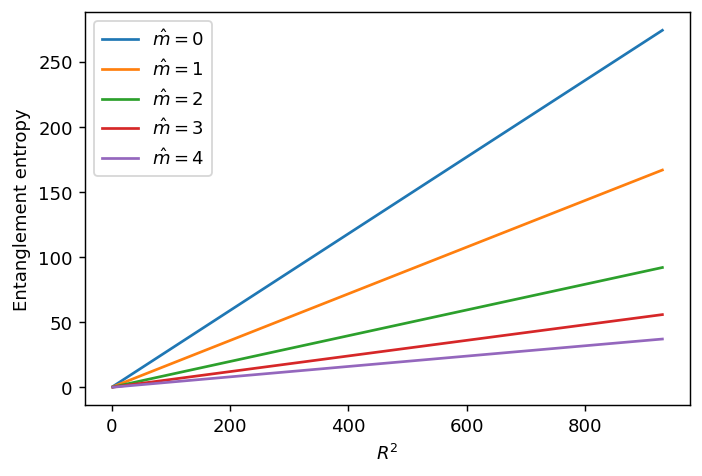}
    \caption{Entanglement entropy vs. $R^2$ for different masses. Addition of mass reduces the entanglement entropy.}
\end{figure}
It is seen that the entanglement entropy decreases with increase in mass, while the linear behaviour with respect to $R^2$ is maintained (Figure \ref{flatspacemass}). The reason for the decrease of the entanglement entropy with increase in mass can be qualitatively understood by the increase in the value of the diagonal terms in the matrix $K$. The off-diagonal terms are indicative of entanglement. If the diagonal terms are much higher in magnitude in comparison, the entanglement entropy decreases. The entropy is now of the form (with $f(0)=1)$)
\begin{equation}
	S=\kappa \,f(\hat{m})\,R^2
\end{equation}
 We tried fits for $f(\hat{m})$ of the forms $\frac{1}{1+p\hat{m}^b}$ and $ce^{-\hat{m}d}$ (Figure \ref{flatmassfit}).
The best-fit results for our entropy curve are: 
\begin{equation}
    S=\kappa\,\frac{r^2}{0.6339 \,\hat{m}^2 +1.6649 }\  ,\hspace{1cm}
    S=\kappa \times 0.9943 \,e^{-0.5013\,\hat{m}} r^2   \ .
\end{equation}

\section{Entanglement Entropy in AdS$_4$}
\subsection{Entanglement Entropy of Scalar Field in AdS$_4$}
Since we are working in curved space now, we have to carefully define the conjugate momentum and the canonical commutation relations.
Let $\mathcal{L}$ be the Lagrangian density and $\pi(x,t)$ be the conjugate momentum. We then make the following definitions (adoped from \cite{Poisson:2009pwt}):
\begin{align}
    \mathcal{L} \ =& \ -\frac{1}{2}g^{\mu \nu}\partial_{\mu} \phi(x) \partial_{\nu} \phi(x)\\
    \pi(x) \ =& \ \frac{\partial(\sqrt{-g}\mathcal{L})}{\partial(\partial_{t}\phi(x))}\\
    \Big[\phi(\vec{x},t),\pi(\vec{y},t)\Big] \ =& \ i\delta^3(\vec{x}-\vec{y})
\end{align}
Now we apply Srednicki's procedure to scalar fields in global AdS$_4$.

\begin{equation}\label{scalarfieldL}
    ds^2 \ = \ -\Big(1+\frac{r^2}{L^2}\Big)dt^2+\frac{dr^2}{\Big(1+\frac{r^2}{L^2}\Big)}+r^2d\theta^2+r^2\text{sin}^2 \theta d\phi^2
\end{equation}
Due to the spherical symmetry of the metric, the solution for the wave equation is once again of the form $\phi_{\ell m}(r)Y_{\ell}^m(\theta,\phi)$. The conjugate momentum is:
\begin{equation}
    \pi=r^2 \text{sin} \theta \frac{1}{1+\frac{r^2}{L^2}}\partial_t \phi
\end{equation}
After integrating out the spherical part and rescaling the field and conjugate momenta as:
\begin{equation}
    \pi_{\ell m} \rightarrow\frac{\sqrt{1+\frac{r^2}{L^2}}}{r}\pi_{\ell m} \hspace{1.5cm} \phi_{\ell m}\rightarrow \frac{r}{\sqrt{1+\frac{r^2}{L^2}}} \phi_{\ell m}
\end{equation}
the Hamiltonian is of the form:
\begin{equation}\label{adsH}
\begin{split}
    H=\frac{1}{2} \sum_{\ell m} \int dr \Bigg[\pi_{\ell m}^2 + r^2\Big(1+\frac{r^2}{L^2}\Big) \Big(\partial_r \Big[\frac{1}{r}\sqrt{1+\frac{r^2}{L^2}}\phi_{\ell m}\Big]\Big)^2
    \\
    +\Big(\frac{\ell(\ell+1)}{r^2}+m^2\Big)\Big(1+\frac{r^2}{L^2}\Big)\phi_{\ell m}^2  \Bigg]
    \end{split}
\end{equation}

\begin{figure}[h]
\centering
\includegraphics[scale=0.5]{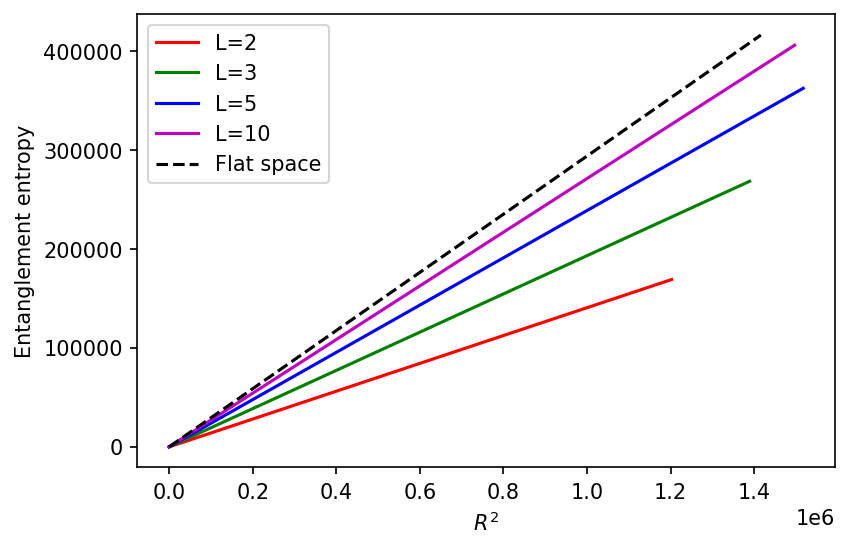}
\caption{Entanglement entropy vs. $R^2$ calculated from working with discretizing proper distance $u$ in radial direction for different values of AdS radius. We see the area law, ie. entanglement entropy $\propto R^2$. We also observe the slope is a function of the AdS radius $L$.}
\label{properdistance}
\end{figure}

\begin{figure}[h]
    \centering
    \includegraphics[scale=0.5]{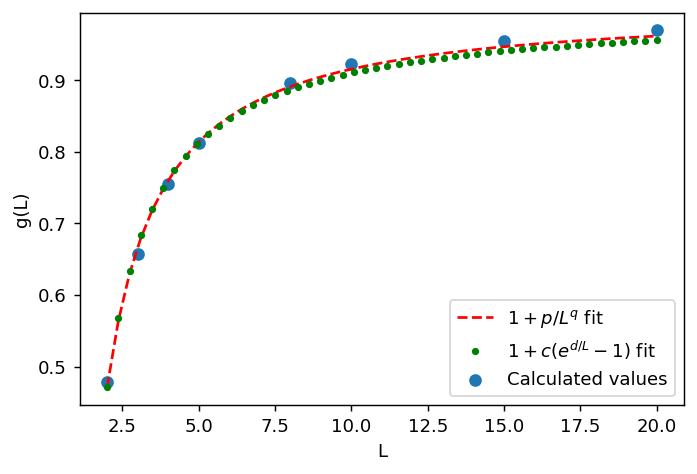}
    \caption{Fits for the function $g(L)$. The functions fitted are $\kappa(1+\frac{p}{L^q})$ with $p=-1.1649,q= 1.1388$ and $\kappa (1+c(e^{d/L}-1))$ with $c=-1.1147,d=0.7762$}
    \label{slope L fit}
\end{figure}

Now we take the proper distance along the $r$ direction to be the discretized variable. The effect of this choice is that the spacing of the divisions is different compared to discretizing $r$. The proper distance $u$ and $r$ are related as follows:
\begin{equation}
\begin{aligned}
du \ =\ &\frac{dr}{\sqrt{1+\frac{r^2}{L^2}}}\\
u\ =\ &L \ \text{sinh}^{-1}\Big(\frac{r}{L}\Big)
\end{aligned}
\end{equation}
We rewrite the Hamiltonian in equation \eqref{adsH} in terms of $u$:
\begin{multline}
    H=\frac{1}{2}\sum_{\ell m} \int du \Bigg[\frac{\text{cosh}^3(\frac{u}{L})}{L^2 \text{sinh}^2(\frac{u}{L})}\pi_{\ell m}^2+L^2 \text{sinh}^2\Big(\frac{u}{L}\Big)\text{cosh}\Big(\frac{u}{L}\Big)(\partial_u \phi_{\ell m})^2 \\ +\Big(\frac{\ell(\ell+1)}{L^2 \text{sinh}^2(\frac{u}{L})}+m^2\Big)L^2 \text{sinh}^2\Big(\frac{u}{L}\Big) \text{cosh}\Big(\frac{u}{L}\Big)\phi_{\ell m}^2\Bigg]
\end{multline}
The rescaling of $\pi_{lm}$ and $\phi_{lm}$ are:
\begin{equation}
    \pi_{\ell m}\rightarrow \frac{\text{cosh}^{3/2}(\frac{u}{L})}{L\text{sinh}(\frac{u}{L})} \pi_{\ell m}\hspace{1.5cm} \phi_{\ell m}\rightarrow \frac{L\text{sinh}(\frac{u}{L})}{\text{cosh}^{3/2}(\frac{u}{L})} \phi_{\ell m}
\end{equation}
\begin{equation}
\begin{split}
H=\frac{1}{2}\sum_{\ell m} \int du \Bigg[ \pi_{\ell m}^2 +  L^2 \text{sinh}^2\Big(\frac{u}{L}\Big) \text{cosh}\Big( \frac{u}{L} \Big) \Bigg( \partial_u \Bigg\{ \frac{\text{cosh}^{3/2}(\frac{u}{L})}{L\text{sinh}(\frac{u}{L})}   \phi_{\ell m}\Bigg\} \Bigg)^2 \\ + \frac{\ell(\ell+1)}{L^2 \ \text{sinh}^2(\frac{u}{L})}\text{cosh}^4 \Big(\frac{u}{L}\Big) \phi_{\ell m}^2 \Bigg]
\end{split}
\end{equation}
We discretize $u$ and compute the entanglement entropy for a given value of $u$ till which we integrate out. The eigenvalue $\xi_{\ell}(n)$ here for large values of $\ell$ is of the form $\frac{L^4 \ \text{sinh}^4 (n/L)}{\ell^2 (\ell+1)^2}$. Since the numerator increases much more rapidly than in the flat space case with increase in $n$, we need to push the summation much further in $\ell$ to obtain good values. The entanglement entropy is found to be proportional to $R^2=L^2 \text{sinh}^2(\frac{u}{L})$, with the slope depending on the AdS radius (Figure \ref{properdistance}).  For increasing values of AdS radius, the slope rapidly approaches the flat space value, which is also expected as we should obtain the flat space results as $L \rightarrow \infty$. The dependence of $L$ on the slope is most visible for very small values of $L$.
The entanglement entropy now is of the form
\begin{equation}
S(r)=\kappa\, g(L)\ R^2,
\end{equation}
where $\kappa$ is the flat space slope and $g(L)$ is a function of $L$ such that $\lim_{L\rightarrow\infty}g(L)=1$.
To study the dependence of the slope on $L$, we tested two functions $g(L)$ of the forms: $1+\frac{p}{L^q}$ and $1+c(e^{d/L}-1)$. Fitting the data to determine the  parameters, we obtained the values $p=-1.1649$, $q=1.1388$ for the first function and $c=-1.1147$,  $d=0.7762$ for the second function(Figure \ref{slope L fit}).

\subsection{Entanglement Entropy of Gauge Fields in AdS$_4$}
In global coordinates, the Lagrangian density is
\begin{equation}
\begin{aligned}
   \sqrt{-g} \mathcal{L}=&-\frac{1}{4} r^2 \sin \theta F_{\mu \nu} F^{\mu \nu} \\
    =& -\frac{1}{4} \ r^2 \sin \theta\, (\partial_{\mu} A_{\nu}-\partial_{\nu} A_{\mu})(\partial^{\mu} A^{\nu}-\partial^{\nu} A^{\mu})
\end{aligned}
\end{equation}
The conjugate momentum is
\begin{equation}
\begin{aligned}
    \pi^{i}=&\frac{\partial}{\partial(\partial_0 A_{i})} (\sqrt{-g}\mathcal{L})\\
    =&-\sqrt{-g}F^{0i},
\end{aligned}
\end{equation}
and we have the following canonical commutation relations:
\begin{equation}
    [\pi^i({\bf{x}}),A_j({\bf{x'}})]=i\delta^i_j \, \delta^3({\bf{x}-{\bf{x'}}})
\end{equation}
If we expand the gauge field and momentum in terms of scalar spherical harmonics and integrate out the angular coordinates, we have:
\begin{equation}
    [\pi^i_{\ell m}(r),A_{j \ell'm'}(r')]=i\delta_{ij}\delta(r-r')\delta_{\ell \ell'}\delta_{mm'}
\end{equation}
Next we write down the Hamiltonian:
\begin{equation}
\begin{aligned}
    \mathcal{H}=& \int drd\theta d\phi \Bigg[-\sqrt{-g}  F^{0i}\partial_0 A_i+\frac{1}{2}\sqrt{-g}(F^{0i}F_{0i}+\tfrac12 F^{ij}F_{ij})\Bigg]\\
    =&\int drd\theta d\phi\frac{-\sqrt{-g}}{2}(F_{0i}F^{0i}-\tfrac{1}{2}F_{ij}F^{ij})\\
    =&-\int dr \frac{r^2}{2} (F_{0i\, \ell m}F^{0i}_{\ell m}-F_{ij\, \ell m}F^{ij}_{\ell m}) 
\end{aligned}
\end{equation}
The indices in the equation take the values $r$, $\theta$ and $\phi$. The calculation of gauge fields in flat space was performed in \cite{Casini:2015dsg} using vector spherical harmonics. We extend this calculation to AdS$_4$ for a spherical entangling surface. Now we define vector spherical harmonics:
\begin{align}
    \bf{Y}_{\ell m}\ =\ &Y_{\ell m} \bf{\hat{r}}  \\
    \bf{\Psi}_{\ell m}\ =\ & \frac{r\nabla Y_{\ell m}}{\sqrt{\ell(\ell+1)}} \label{vshpsi}\\
    \bf{\Phi}_{\ell m}\ =\ & \bf{\hat{r}} \times \bf{\Psi}_{\ell m} \label{vshphi}
\end{align}
Using the expression for gradient in spherical polar coordinates, \eqref{vshpsi} and \eqref{vshphi} can be rewritten as:
\begin{align}
    {\bf {\Psi}_{\ell m}}=&\frac{1}{\sqrt{\ell(\ell+1)}} \Big(\partial_{\theta}Y_{\ell m}\, \hat{\theta}+\frac{1}{\sin \theta}\partial_{\phi}Y_{\ell m}\, \hat{\phi}\Big) \\
    {\bf {\Phi}_{\ell m}}=&\frac{1}{\sqrt{\ell(\ell+1)}} \Big(\partial_{\theta}Y_{\ell m}\, \hat{\phi}-\frac{1}{\sin \theta}\partial_{\phi}Y_{\ell m}\, \hat{\theta}\Big)
\end{align}
Vector spherical harmonics constitute an orthonormal basis:
\begin{equation}
    \int {\bf Y_{\ell m}}.{\bf Y^*_{\ell' m'}}d\Omega=\int {\bf Y_{\ell m}}.{\bf Y^*_{\ell' m'}}d\Omega= \int {\bf Y_{\ell m}}.{\bf Y^*_{\ell' m'}}d\Omega= \delta_{\ell \ell'} \delta_{m m'}
\end{equation}
while all other combinations give zero. A vector field can be expanded using the above basis as:
\begin{equation}\label{svh}
    \textbf{V}=\sum_{l=0}^{\infty} \sum_{m=-l}^{\ell} \Bigg( V^r_{\ell m} (r) {\bf Y}_{\bf{\ell m}} + V^{(1)}_{\ell m}(r) {\bf{\Psi}_{\bf{\ell m}}} + V^{(2)}_{\ell m}(r) \bf{\Phi}_{\bf{\ell m}} \Bigg)
\end{equation}
We have to first define how electric and magnetic fields are defined covariantly in a background where the $g_{tt}$ component is non zero. The electric field is defined as:
\begin{equation}
    E^i=\sqrt{-g_{tt}}F^{0i}
\end{equation}
To define the magnetic field, we make use of the four-dimensional Levi-Civita tensor:
\begin{equation}
    B^i=\frac{1}{2}\epsilon^{i0jk}\sqrt{g_{tt}}F_{jk}
\end{equation}
where $\epsilon^{\mu \nu \rho \sigma}$ is the four dimensional Levi-Civita tensr density. The above equation can be shown to be equal to:
\begin{equation}
    B^i=-\frac{1}{2\sqrt{h}}\varepsilon^{ijk}F_{jk},
\end{equation}
where $h$ is the determinant of the spatial part of the metric and $\varepsilon^{ijk}$ is the three-dimensional Levi-Civita symbol. Since vector spherical harmonics define an orthornormal basis, we have to define the basis vectors in our coordinates and vector components accordingly. The orthonormal basis vectors in ($r,\theta, \phi$) coordinates are:
\begin{equation}
    \hat{r}=\sqrt{1+\tfrac{r^2}{L^2}} \begin{pmatrix}
        1 \\
        0 \\
        0
    \end{pmatrix} 
    \hspace{1.5cm}
    \hat{\theta}=\frac{1}{r} \begin{pmatrix}
        0 \\
        1 \\
        0
    \end{pmatrix}
    \hspace{1.5cm}
    \hat{\theta}=\frac{1}{r \sin \theta} \begin{pmatrix}
        0 \\
        1 \\
        0
    \end{pmatrix}
\end{equation}
The dot products of these vectors have the property that $\hat{r}\cdot\hat{r}=\hat{\theta}\cdot\hat{\theta}=\hat{\phi}\cdot\hat{\phi}=1$ in the metric \eqref{scalarfieldL} and zero otherwise. 
The contravariant vector components will pick up the inverse of these factors. This means that the orthogonal vector components $E^i_O$ of the electric field are:
\begin{equation}
    E^r_O=F^{0i} \hspace{1.5cm} E^{\theta}_O=\sqrt{1+\tfrac{r^2}{L^2}}F^{0\theta}r \hspace{1.5cm} E^{\phi}_O=\sqrt{1+\tfrac{r^2}{L^2}}F^{0\phi}r \sin \theta
\end{equation}
Similarly, the components of the magnetic field in the orthogonal basis are:
\begin{equation}
    B^r_O=-\frac{F^{\theta \phi}}{r^2 \sin \theta} \hspace{1.5cm} B^{\theta}_O=-\frac{F^{r \phi}}{r \sin \theta} \hspace{1.5cm} B^{\phi}_O=-\frac{F^{r \theta}}{r},
\end{equation}
In global coordinates, the analogue of $\nabla \cdot \textbf{E}=0$ is:
\begin{equation}
    \partial_{i}(\sqrt{-g}F^{0 i})=0.
\end{equation}
By making use of the relation between $F^{0i}$ and the contravariant components $E^i_O$ of the electric field in the orthonormal basis, we have:
\begin{equation}
    \partial_r (r^2 \sin \theta E^r_O)+\partial_{\theta} \Bigg(r^2 \sin \theta \frac{E^{\theta}_O}{r \sqrt{1+\tfrac{r^2}{L^2}}}\Bigg)+\partial_{\phi} \Bigg(r^2 \sin \theta \frac{E^{\theta}_O}{r \sin \theta\sqrt{1+\tfrac{r^2}{L^2}}}\Bigg)
\end{equation}

\begin{equation}
    \frac{1}{r^2} \sqrt{1+\frac{r^2}{L^2}} \partial_r ( r^2 E^r_O)+\frac{1}{r\, \text{sin} \theta} \partial_{\theta}(\text{sin}\theta E^{\theta}_O)+\frac{1}{r\, \text{sin} \theta} \partial_{\phi}  E^{\phi}_=0
\end{equation}
In terms of vector spherical harmonics \eqref{svh}:
\begin{equation}
    \sqrt{1+\frac{r^2}{L^2}}\partial_r E^r_{\ell m} + \frac{2}{r}E^r_{\ell m}\sqrt{1+\frac{r^2}{L^2}}-E^{(1)}_{\ell m} \frac{\sqrt{\ell(\ell+1)}}{r}=0
\end{equation}
For $l=0$, $E^r_{\ell=0}(r)\sim \frac{1}{r^2}$. In the absence of charges, the only solution that is finite at the boundary is $E^r_{l=0}=0$. Similar to the case in flat space\cite{Casini:2015dsg}, $E^{(1)}_{\ell m}$ can be written in terms of $E^{r}_{\ell m}$. The same follows for $\textbf{B}$ because of $\nabla \cdot \textbf{B}=0$. The Hamiltonian is
\begin{equation}\label{ads4gaugeH}
\begin{aligned}
    H=&\frac{1}{2}\int dr d\theta d\phi \sqrt{-g}\, (\textbf{E}^2+\textbf{B}^2)\\
    =& \frac{1}{2} \sum_{\ell=1}^{\infty} \sum_{m=-\ell}^\ell \int dr \, r^2 \Bigg[(E^r_{\ell m})^2 + (B^{(2)}_{\ell m})^2 +\Bigg(\sqrt{1+\frac{r^2}{L^2}}\frac{r}{\sqrt{\ell(\ell+1)}}\partial_r E^r_{\ell m}\\
    & + \frac{2}{\sqrt{\ell(\ell+1)}}E^r_{\ell m} \sqrt{1+\frac{r^2}{L^2}}\Bigg)^2\Bigg]+(E_{\ell m} \leftrightarrow B_{\ell m}),
\end{aligned}
\end{equation}
where the summation over $\ell$ starts from 1. Next, we need the commutation relations between components of {\bf E} and {\bf B} in vector spherical harmonics so that we can appropriately rescale them. We first derive the commutation relations in spherical polar coordinates and use them to obtain the commutation relations in vector spherical harmonics. We illustrate this for $E^r_O$ and $B^{\theta}_O$:
\begin{equation}
\begin{aligned}
    [E^r_O({\bf{r}}),B^{\theta}_O({\bf{r'}})]=&[F^{0r}({\bf{r}}),-F^{r\phi}({\bf{r'}})\tfrac{1}{r' \sin \theta'}\sqrt{1+\tfrac{r'^2}{L^2}}] \\
    =& [F^{0r}({\bf{r}}),\partial_{\phi}A_r({\bf{r'}})\tfrac{1}{r' \sin \theta'}\sqrt{1+\tfrac{r'^2}{L^2}}]
\end{aligned}
\end{equation}
In going to the second line, we have used the fact that the other term in $F^{r \phi}$ does not contribute to the commutation relation. We can further expand $F^{0r}$ and $A_r$ in terms of scalar spherical harmonics:
\begin{align}
    F^{0r}({\bf{r}})=&\sum_{\ell m}F^{0r}_{\ell m}(r)Y_{\ell m}(\theta,\phi)\\
    A_r({\bf{r}})=&\sum_{\ell m}A_{r\,\ell m}(r)Y_{\ell m}(\theta,\phi),
\end{align}
We integrate over $\theta$ and $\phi$ and simplify to an expression in terms of the scalar spherical harmonics:
\begin{equation}
    [E^r_{O,\ell m}({\bf{r}}),B^{\theta}_{O,\ell' m'}({\bf{r'}})]=\sum_{\ell m}\frac{-i\sqrt{1+\tfrac{r^2}{L^2}}}{r^3 \sin \theta} Y_{\ell m}(\theta,\phi) \partial_{\phi}Y_{\ell m}(\theta',\phi')\delta (r-r')
\end{equation}
where we have made use of the commutation relation:
\begin{equation}
    [r^2 F^{0r}_{\ell m}(r),A_{r\, \ell'm'}(r')]=i\delta(r-r')\delta_{\ell \ell'} \delta_{mm'}
\end{equation}
The $r^2$ in the first term with $F^{0r}$ appears due to $\sqrt{-g}$ containing an $r^2$, which gets included in the definition of the conjugate momentum. Similarly, we can also see that:
\begin{equation}
    [E^r_{O,\ell m}({\bf r}),B^{\phi}_{O,\ell' m'}({\bf r'})]=\sum_{\ell m}\frac{i \sqrt{1+\tfrac{r^2}{L^2}}}{r^3}Y_{\ell m}(\theta,\phi) \partial_{\theta}Y_{\ell m}(\theta',\phi') \delta(r-r')
\end{equation}

Now we know that
\begin{equation}
\begin{aligned}
    B^{(1)}_{\ell m}(r)=&\int {\bf B}.{\bf \Psi}^*_{\ell m} d\Omega  \\
    =& \int \sum_{\ell' m'} \Big(B^{\theta}_{O,\ell' m'}(r)\hat{\theta}+B^{\phi}_{O,\ell' m'}(r)\hat{\phi}\Big)Y_{\ell' m'}(\theta,\phi)\cdot\frac{1}{\sqrt{\ell (\ell+1)}} \Bigg(\frac{\partial Y^*_{\ell m}}{\partial \theta} \hat{\theta}+\frac{1}{\sin \theta} \frac{\partial Y^*_{\ell m}}{\partial \phi} \hat{\phi}\Bigg) \\
    =& \sum_{\ell' m'} \int \frac{d\Omega}{\sqrt{\ell (\ell+1)}} Y_{\ell' m'}(\theta,\phi) \Bigg[B^{\theta}_{O,\ell' m'} (r)\frac{\partial Y^*_{\ell m}}{\partial \theta}+B^{\phi}_{O,\ell' m'}(r)\frac{1}{\sin \theta}\frac{\partial Y^*_{\ell m}}{\partial \phi}\Bigg]
\end{aligned}
\end{equation}

Similarly,
\begin{equation}
    B^{(1)}_{\ell m}(r)=\sum_{\ell' m'} \int \frac{d\Omega}{\sqrt{\ell (\ell+1)}} Y_{\ell' m'}(\theta,\phi) \Bigg[-B^{\theta}_{O,\ell' m'} (r)\frac{1}{\sin \theta}\frac{\partial Y^*_{\ell m}}{\partial \phi}+B^{\phi}_{O,\ell' m'}(r)\frac{\partial Y^*_{\ell m}}{\partial \theta}\Bigg]
\end{equation}

Now, we can use the expressions above to write the commutator of the components in vector spherical harmonics.
\begin{multline}
    [E^r_{\ell m}(r),B^{(2)}_{\ell' m'}(r')]=\sum_{\tilde{\ell}\tilde{m}} \int \frac{d\Omega}{\sqrt{\ell (\ell+1)}}Y_{\tilde{\ell}\tilde{m}}(\theta,\phi) \Bigg(-[E^r_{O,\ell m}(r),B^{\theta}_{O,\ell m}(r')]\frac{1}{\sin \theta} \frac{\partial Y^*_{\ell' m'}}{\partial \phi}\\
    +[E^r_{O,\ell m}(r),B^{\theta}_{O,\ell m}(r')] \frac{\partial Y^*_{\ell' m'}}{\partial \theta}\Bigg)
\end{multline}
Substituting the commutators, we have:
\begin{multline}
     [E^r_{\ell m}(r),B^{(2)}_{\ell' m'}(r')]= \sum_{\tilde{l}\tilde{m}} \int \frac{d\Omega d\Omega' \sqrt{1+\tfrac{r^2}{L^2}}}  {\sqrt{\ell(\ell+1)}} \Bigg(\frac{Y_{\tilde{\ell} \tilde{m}}(\theta,\phi)}{\sin \theta} \frac{\partial Y^*_{\ell' m'}(\theta,\phi)}{\partial \phi} Y^*_{\tilde{\ell} \tilde{m}}(\theta',\phi') \frac{i}{r^3 \sin \theta'} \partial_{\phi}Y_{\ell m} (\theta',\phi')\\
    + Y_{\tilde{l}\tilde{m}}(\theta,\phi)\partial_{\phi}Y^*_{\ell' m'}(\theta,\phi)\frac{i Y_{\tilde{\ell}\tilde{m}}(\theta',\phi')}{r^3} \partial_{\theta}Y_{\ell m}(\theta',\phi') \Bigg)
\end{multline}
Finally, using the orthogonality of scalar spherical harmonics:
\begin{equation}
    \sum_{\ell m} Y_{\ell m}(\theta,\phi) Y^*_{\ell m}(\theta',\phi')=\delta(\cos \theta-\cos \theta')\delta(\phi-\phi'),
\end{equation}
we simplify the commutation relation:
\begin{equation}
\begin{aligned}
    [E^r_{\ell m}(r),B^{(2)}_{\ell' m'}(r')]=&\frac{d\Omega\,i\sqrt{1+\tfrac{r^2}{L^2}}}{r^3\sqrt{\ell (\ell+1) }}\Bigg(\frac{1}{\sin ^2 \theta} \partial_{\phi}Y^*_{\ell m}(\theta,\phi)\partial_{\phi}Y_{\ell' m'}(\theta,\phi)+\partial_{\theta}Y^*_{\ell m}(\theta,\phi)\partial_{\theta}Y_{\ell'm'}(\theta,\phi)\Bigg)\\
    =&i\frac{\sqrt{\ell(\ell+1)(1+\tfrac{r^2}{L^2}})}{r^3} \delta_{\ell \ell'} \delta_{m m'}
\end{aligned}
\end{equation}
We can also find that:
\begin{equation}
     [B^r_{\ell m}(r),E^{(2)}_{\ell' m'}(r')]=-i\frac{\sqrt{\ell(\ell+1)(1+\tfrac{r^2}{L^2}})}{r^3} \delta_{\ell \ell'} \delta_{m m'}
\end{equation}
Now that we have obtained our commutation relations, we can rescale the conjugate variables:
\begin{equation}
\begin{aligned}
    &B^{(2)}_{\ell m}\rightarrow r B^{(2)}_{\ell m} \hspace{2.5cm} E^r_{\ell m} \rightarrow \frac{r^2 E^r_{\ell m}}{\sqrt{\ell(\ell+1)(1+\tfrac{r^2}{L^2})}}\\
    &B^{(2)}_{\ell m}\rightarrow r B^{(2)}_{\ell m} \hspace{2.5cm} E^r_{\ell m} \rightarrow \frac{r^2 E^r_{\ell m}}{\sqrt{\ell(\ell+1)(1+\tfrac{r^2}{L^2})}},
\end{aligned}
\end{equation}
and rewrite the Hamiltonian in equation \eqref{ads4gaugeH} as:
\begin{multline}
    H=\frac{1}{2} \sum_{l=1}^{\infty}\sum_{m=-l}^l \int dr\Bigg[(B^{(2)}_{lm})^2+\frac{l(l+1)(1+\frac{r^2}{L^2})}{r^2 }(E^r_{lm})^2 +\Big( 1+\frac{r^2}{L^2} \Big) \Big(\partial_r \Big[E^r_{\ell m}\sqrt{1+\tfrac{r^2}{L^2}}\Big]\Big)^2 \Bigg]  \\
     + (E_{lm}\leftrightarrow B_{lm}).
\end{multline}
We then change the variable to the proper distance along radial direction $s$:
\begin{multline}\label{entgaugeads}
    H=\, \frac{1}{2} \sum_{\ell=1}^{\infty}\sum_{m=-\ell}^{\ell} \int du  \Bigg[(B^{(2)}_{\ell m})^2 + \Big\{\partial_u \Big(E^r_{\ell m}\cosh^{3/2}(\tfrac{u}{L})\Big) \Big\}^2 + \frac{\ell(\ell+1)\cosh ^4(\tfrac{u}{L})}{L^2 \text{sinh}^2(\frac{u}{L})} (E^r_{\ell m})^2 \Bigg]  \\
     + (E_{\ell m}\leftrightarrow B_{\ell m})
\end{multline} 
The steps to discretize the Hamiltonian are exactly the same as for the cases discussed before. Since we see that there are two copies of the harmonic-oscillator-like Hamiltonian in equation \eqref{entgaugeads}, the entanglement entropy is twice the answer we get from the numerical calculation. Also, in this case, we do not see much variation with respect to different values of $L$.
\begin{figure}[h]
    \centering
    \includegraphics[scale=0.7]{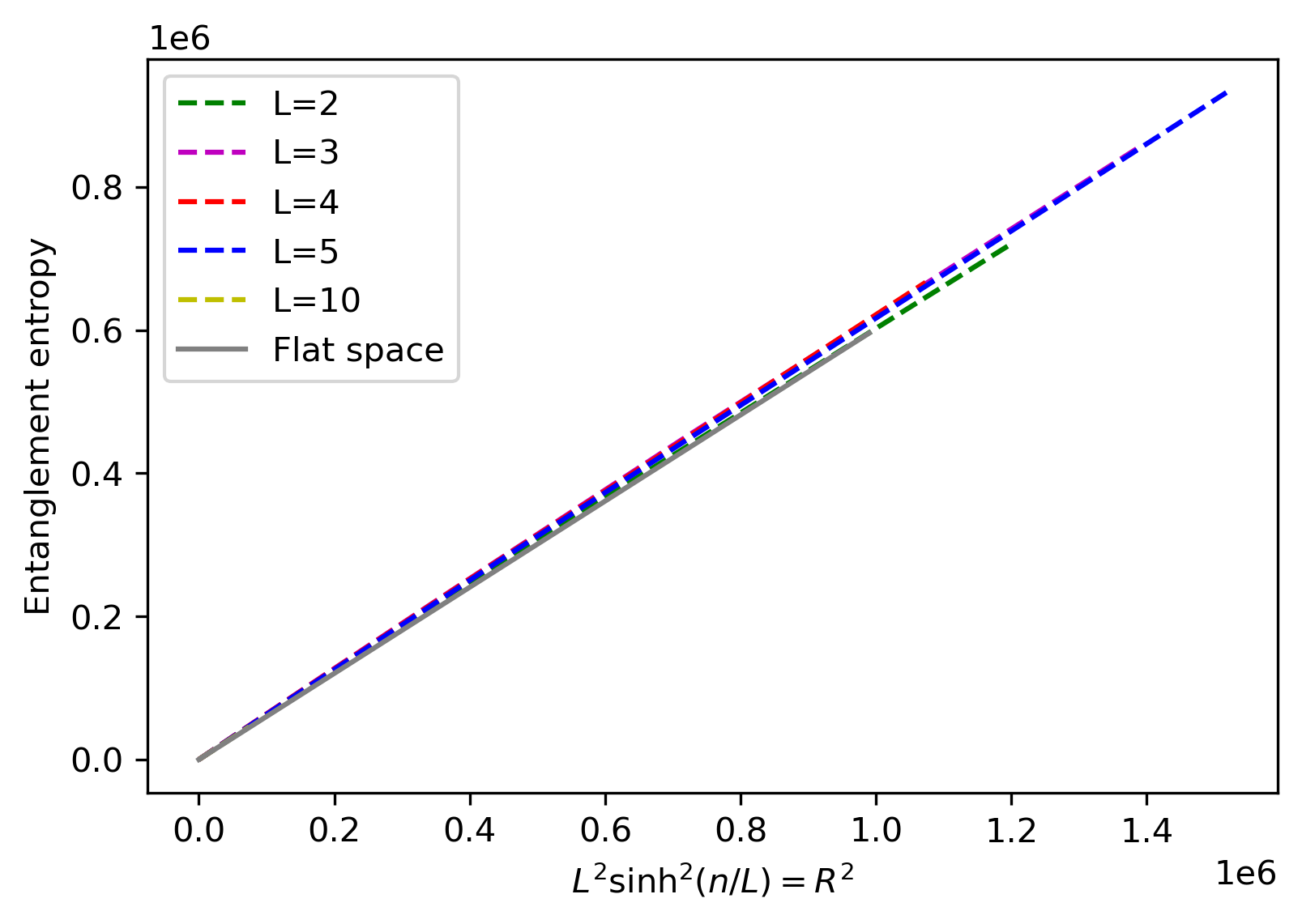}
    \caption{Entanglement entropy of gauge fields in global AdS$_4$ for different values of $L$.}
\end{figure}
\section{Entanglement Entropy of Fields inside RT Surface}
\subsection{Entanglement Entropy of Scalar Field within the RT surface}
The Klein Gordon equation in curved space is
\begin{equation}
	\Bigg[\frac{1}{\sqrt{-g}}\partial_{\mu}(\sqrt{-g} g^{\mu \nu} \partial_{\nu}) + m^2\Bigg] \phi=0.
\end{equation}
We look for a solution in terms of the coordinates $\tilde{t},\tilde{\eta}$ and $\tilde{x}$ in the metric \eqref{rtfoliation}.
Assuming a variable separable form for the solution:
\begin{equation}
    \phi(\tilde{\eta},\tilde{x},\tilde{t})=e^{i \omega t}  A(\tilde{\eta})B(\tilde{x}),
\end{equation}
the equation of motion is
\begin{equation}
\begin{split}
    \frac{A''(\tilde{\eta})+\tanh (\tilde{\eta}) A'(\tilde{\eta})}{A(\tilde{\eta})} + \omega^2 \text{sech}^2(\tilde{\eta})\hspace{2in} \\ + \frac{\cosh ^2(\tilde{x}) B''(\tilde{x})+\sinh (2 \tilde{x}) B'(\tilde{x})-L^2 m^2 B(\tilde{x}) \cosh ^2(\tilde{x})}{B(\tilde{x})} 
    =0.
    \end{split}
\end{equation}
Solving for the $\tilde{x}$ part alone:
\begin{equation}
    \frac{\cosh ^2(\tilde{x}) B''(\tilde{x})+\sinh (2 \tilde{x}) B'(\tilde{x})-L^2 m^2 B(\tilde{x}) \cosh ^2(\tilde{x})}{B(\tilde{x})}=-q,
\end{equation}
we find solutions in terms of associated Legendre polynomials.
\begin{equation}
    B(\tilde{x})= c_1\ \text{sech} \tilde{x}\ P_{\ell}^{m'}(\tanh \tilde{x})+c_2\ \text{sech} \tilde{x}\ Q_{\ell}^{m'}(\tanh \tilde{x}),
\end{equation}
where $\ell=\frac{1}{2}(\sqrt{1+4q}-1)$ and $m'=\sqrt{L^2 m^2+1}$. We discard the second term, since it blows up at $x=0$, and take only the first term, where $\tilde{x}$ takes values from $-\infty$ to $\infty$. For normalizable wave function, we need $\ell$ to take values over $\mathbb{N}$, which makes $q$ take values  $\ell(\ell+1)$. For the massless case, this gives us the solution  of the form $\text{sech}\tilde{x}\ P_{\ell}^1(\text{tanh}\tilde{x})$. The scalar field is now of the form 
\begin{equation}
   \phi=\sum_l \phi_l(\tilde{\eta})\, P_l^1(\tanh \tilde {x}) 
\end{equation}
 The conjugate momentum for $\phi$ is:
\begin{equation}
\begin{aligned}
    \pi& =\sqrt{-g}g^{\tilde{t}\tilde{t}} \partial_{\tilde{t}}\phi \\
    & = L \frac{\partial_{\tilde{t}}\phi}{\text{cosh} \tilde{\eta}} 
\end{aligned}
\end{equation}
Now the Hamiltonian density can be written down:
\begin{equation}
    \mathcal{H}=\frac{1}{2L}\cosh(\tfrac{\eta}{L}) \pi^2+\frac{1}{2L} \cosh(\tfrac{\eta}{L})(\partial_{\eta} \phi)^2 +L\cosh(\tfrac{\eta}{L}) \cosh^2(\tfrac{x}{L})(\partial_x \phi)^2
\end{equation}
Substituting the solutions of $\phi$ and $\pi$ and integrating out $x$, we obtain the Hamiltonian:
\begin{equation}
\begin{aligned}
    H=&\int dx\, d\eta \,  \mathcal{H}\\
    =&\frac{1}{2}\sum_{\ell}\int d\eta \Bigg[ \frac{1}{L}\cosh(\tfrac{\eta}{L}) \pi_{\ell}^2+L \cosh(\tfrac{\eta}{L}) (\partial_{\eta} \phi_{\ell})^2) +L\cosh(\tfrac{\eta}{L})\ell(\ell+1)\phi_{\ell}^2\Bigg],
\end{aligned}
\end{equation}
where $\ell$ takes values $1,2,3...$. and so on. We discretize the $\eta$ coordinate and rescale the field and conjugate momentum as follows:
After rescaling the field and momentum:
\begin{equation}
    \pi_{\ell} \rightarrow \sqrt{\frac{\cosh(\tfrac{\eta}{L})}{L}} \pi_{\ell} \hspace{1.5cm} \phi_{\ell}  \rightarrow \sqrt{\frac{L}{\cosh(\frac{\eta}{L})}} \phi_{\ell}
\end{equation}

\begin{figure}[h]
    \centering
    \includegraphics[scale=0.6]{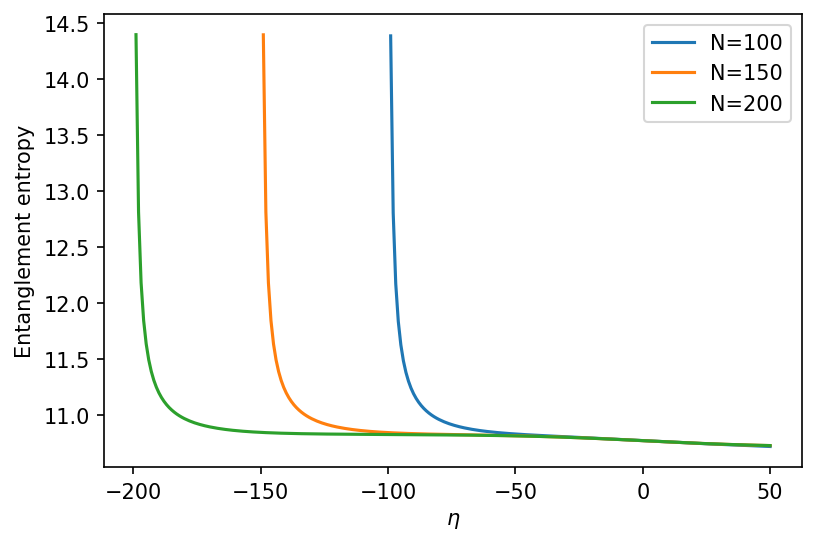}
    \caption{Entanglement entropy vs. discretized values of $\eta$ for $L=50$ where the $\eta=ia$ where i is the index. Values are plotted for different values of $N$. The values agree after the initial fall. This shows that the initial fall is an IR effect near the boundary of AdS.}
    \label{differentN}
\end{figure}

\begin{figure}[h]
    \centering
    \includegraphics[scale=3]{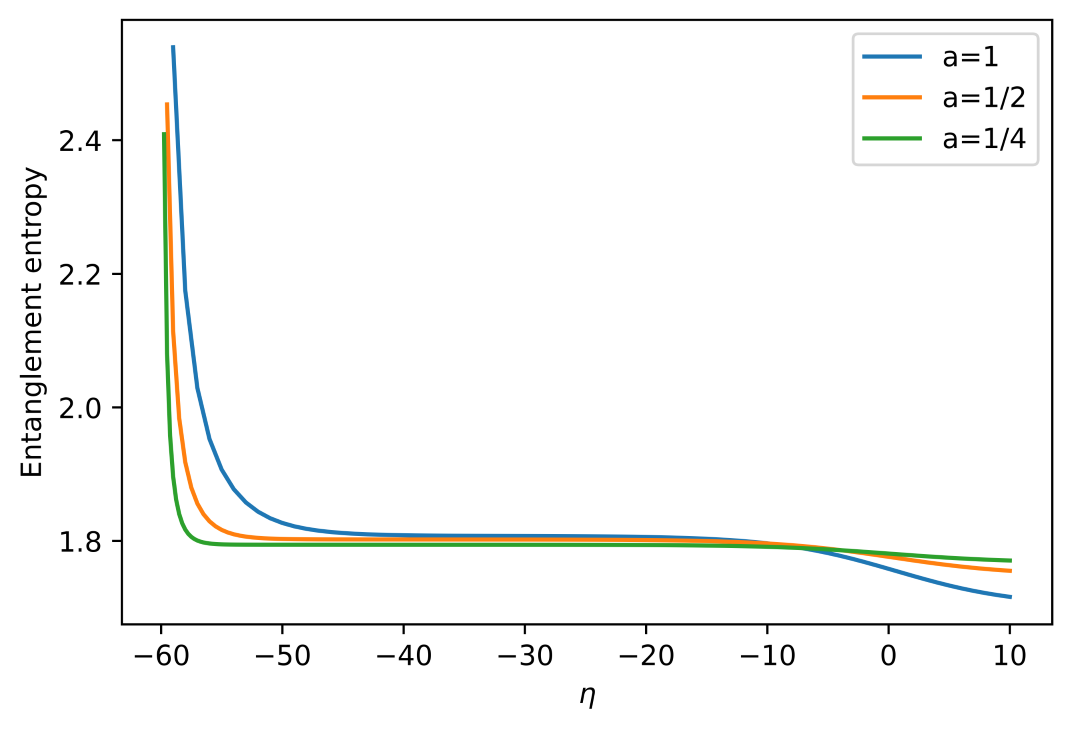}
    \caption{Entanglement entropy vs. $\eta$ for $L=10$ for different values of $a$, where $\eta=ja$ for discrete index $j$. The almost-constant behaviour of entanglement entropy improves with smaller spacing, while the initial steep decrease persists.}
    \label{diffa}
\end{figure}

This gives us the final form of the Hamiltonian:
\begin{equation}
\begin{split}
    H=\sum_{\ell,j} \frac{1}{2a}\Bigg[\pi_{\ell,j}^2 + \text{cosh}\Big(\tfrac{j-N}{L}\Big)\Bigg(\sqrt{\text{cosh} \Big(\tfrac{j+1-N}{L}\Big)}\phi_{\ell,j+1}-\sqrt{\text{cosh} \Big(\tfrac{j-N}{L}\Big)}\phi_{\ell,j}\Bigg)^2 \\
    + \frac{1}{L^2}\text{cosh}^2\Big(\tfrac{j-N}{L}\Big)\, \ell(\ell+1)\phi_{\ell,j}^2 \Bigg].
\end{split}
\end{equation}
The value of $N$ corresponds to the highest value of $\eta$ that we would consider. Given a value of $\eta=\eta_0$, integrating over the fields starting from $\eta_{max}$ (determined by $N$) to $\eta_0$ would correspond to finding the entanglement entropy of fields within the RT surface. We find that the entanglement entropy largely remains constant, after a steep decrease. We perform the calculations for different values of $N$, where a larger value of $N$ amounts to going closer to the boundary (Figure \ref{differentN}).  The initial fall off appears for all the cases, which indicates that it is associated with IR divergences associated with the AdS boundary.
We carry out the computation for different values of $L$ and see that the entanglement entropy increases with increase in $L$. 
\begin{figure}[h]
    \centering
    \includegraphics[scale=1.2]{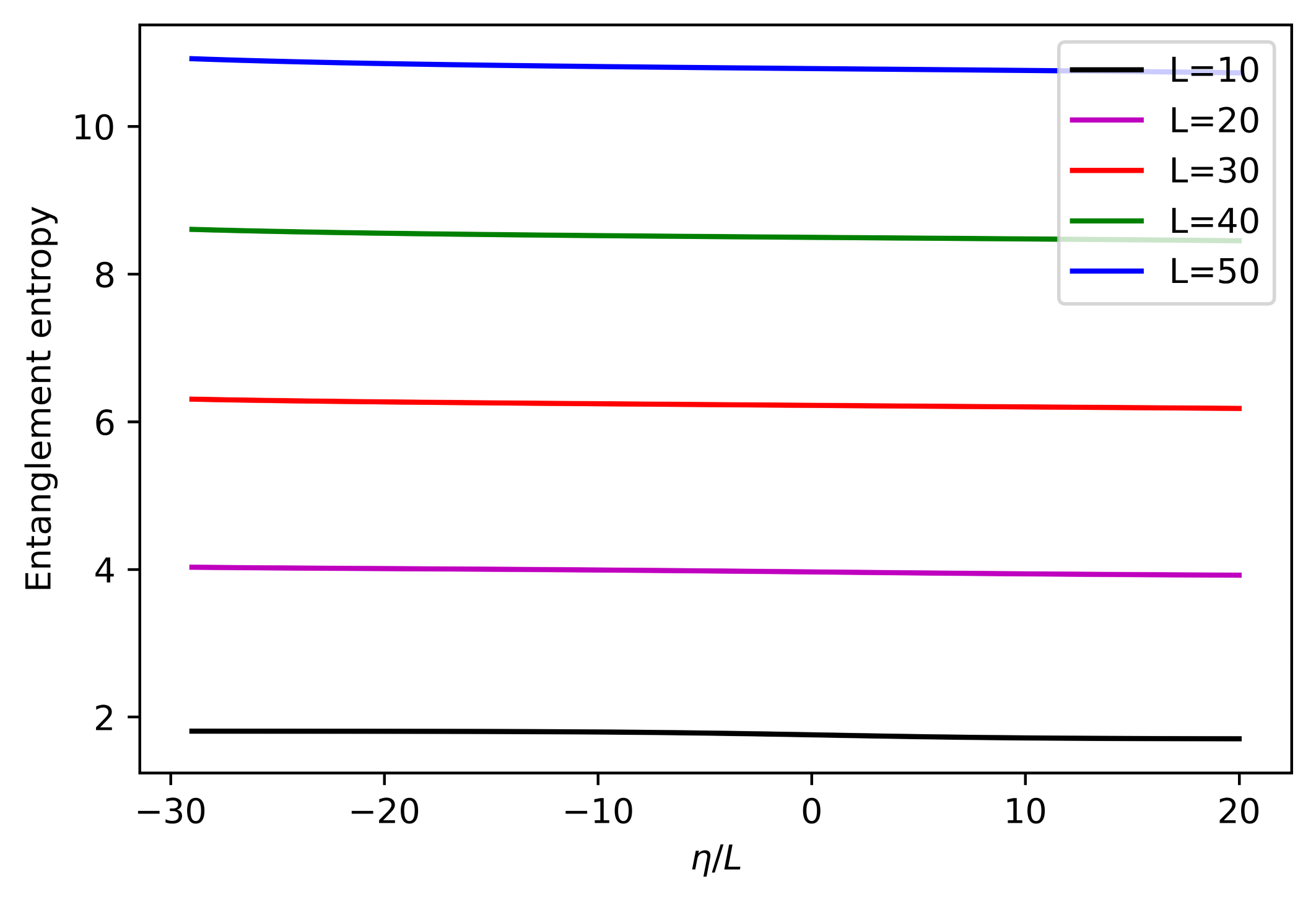}
    \caption{Entanglement entropy for scalar fields inside the RT surface in global AdS$_3$ for different $L$. We plot the range of values after the initial fall off, where the entropy remains almost constant. Entanglement entropy increases with increasing values of $L$.}
\end{figure}
Fitting the data where the entropy is almost constant with a form $p(\frac{L}{a})+q$ gives the values $p=0.226$ and $q=-0.532$. 
Repeating the calculation for different spacing (ie. different values for $a$) seems to make the values better, in that they approach the constant values faster compared to the cases of larger spacing (figure \ref{diffa}). It also shows that the dip in the values around $\eta=0$ for larger values of spacing is an artefact of the numerics.
The metric tells us that this constant behaviour of entanglement entropy as a function of $\eta$ is indeed the expected area law, since for fixed $\eta$, the metric is $L^2 d\tilde{x}^2$. Moreover, the cutoff in $\tilde{x}$ is the same for all constant $\tilde{\eta}$ surfaces. In the derivation of the RT formula in global coordinates, the cutoff is imposed on the $r$, which is different from a cutoff in $x$, and hence in that case, we see that the entanglement entropy will behave differently.

\subsection{Gauge Fields in AdS$_3$ within the RT surface}
To work out the problem of gauge fields, we need to find the suitable basis for the gauge fields in the $\tilde{x}$ coordinate. To keep better track of rescaling the components and raising/lowering of indices, we use vielbeins:
\begin{align}
    g_{\mu\nu}=&e^a_{\mu}e^b_{\nu}\eta_{ab} \\
    E^{\nu}_a=& \eta_{ab} g^{\mu \nu} e^b_{\nu} \\
    E^{\mu}_a e^b_{\mu}=&\delta^b_a
\end{align}
The gauge-invariant Lagrangian is:
\begin{equation}
\begin{aligned}
    \mathcal{L}=&-\frac{1}{4}F\wedge*F \\
    =& -\frac{1}{2}\Big[(F_{01})^2+(F_{02})^2-(F_{12})^2\Big] e^0 \wedge e^1 \wedge e^2
\end{aligned}
\end{equation}
where $F=dA$ is a two-form, $A$ is the one-form gauge field and $*F$ is the Hodge dual. For the metric that we are using, the vielbein components $e^a_{\mu}$ and their inverses $E_{a}^{\mu}$ are:
\begin{align}
    e^0_{\tilde{t}}=& L \text{cosh} \tilde{x} \text{cosh} \tilde{\eta} \hspace{3cm} E^{\tilde{t}}_0=\frac{1}{L} \text{sech} \tilde{x} \text{sech} \tilde{\eta}\\
    e^1_{\tilde{\eta}}=& L \text{cosh} \tilde{x} \hspace{3.9cm}  E^{\tilde{\eta}}_1=\frac{1}{L} \text{sech} \tilde{x}  \\
    e^2_{\tilde{x}}=& L \hspace{4.8cm} E^{\tilde{x}}_2=\frac{1}{L} 
\end{align}
The basis used for the gauge field is:
\begin{align}
    A_0=& \sum_{\ell} A_{0 \ell}(\eta,t)\, \text{sech}\tilde{x} P^1_{\ell}(\text{tanh} \tilde{x}) \\
    A_1=& \sum_{\ell} A_{1 \ell}(\eta,t)\, \text{sech}\tilde{x} P^1_{\ell}(\text{tanh} \tilde{x})
\end{align}
This basis is motivated by the solution for the equation of motion of the scalar field in this background. Such an expansion is allowed because of the completeness of the scalar spherical harmonics\cite{RGBarrera_1985}. The gauge choice we make is $A_2=0$. This proves to be useful because it enables us to integrate out $\tilde{x}$ from the Hamiltonian.
The components of $F$ in the vielbein basis are:
\begin{align}
    F_{01}=& \frac{1}{L}\text{sech}\tilde{x}(\text{sech}\tilde{\eta} \partial_{\tilde{t}}A_1- \partial_{\tilde{\eta}}A_0)\label{f01}\\
    F_{02}=&-\frac{1}{L}\partial_{\tilde{x}}A_0\\
    F_{12}=&-\frac{1}{L}\partial_{\tilde{x}}A_1 \label{f12}
\end{align}
The Lagrangian is
\begin{equation}
\begin{aligned}\label{gaugeL}
    \mathcal{L}=-\frac{1}{4}\int F \wedge *F = & -\frac{1}{2}\int \Big[(F_{01})^2+(F_{02})^2-(F_{12})^2\Big]e^0\wedge e^1 \wedge e^2\\
    =&-\frac{1}{2}\int \sqrt{-g} \Big[(F_{01})^2+(F_{02})^2-(F_{12})^2\Big] d\tilde{t}d\tilde{\eta}d\tilde{x} 
\end{aligned}
\end{equation}
Substituting \eqref{f01}-\eqref{f12} into \eqref{gaugeL} and integrating out $\tilde{x}$, we have the Lagrangian that depends only on $\eta$: 
\begin{multline}
    \mathcal{L}=\frac{1}{2}\sum_{\ell} \int d\tilde{\eta} \Big[\text{sech}\tilde{\eta}(\partial_{\tilde{t}} A_{1 \ell})^2 + \text{cosh} \tilde{\eta} (\partial_{\tilde{\eta}} A_{0 \ell})^2-2(\partial_{\tilde{t}} A_{1 \ell})(\partial_{\tilde{\eta}} A_{0 \ell}) \\ + \text{cosh} \tilde{\eta} A_{0l}^2 {\ell}(\ell+1)-\text{cosh} \tilde{\eta} A_{1l}^2 \ell(\ell+1) \Big]
\end{multline}
$A_{0 \ell}$ can be eliminated using the Gauss law constraint:
\begin{equation}
    \partial_{\tilde{\eta}}(\text{cosh} \tilde{\eta} \partial_{\tilde{\eta}} A_{0 \ell})-\partial_{\tilde{\eta}} \partial_{\tilde{t}}A_{1 \ell}-\text{cosh} \tilde{\eta} A_{0 \ell} \ell(\ell+1)=0
\end{equation} 
The conjugate momentum corresponding to $A_{1 \ell}$ is:
\begin{equation}
    \pi_{1 \ell}= \text{sech} \tilde{\eta} (\partial_{\tilde{t}}A_{1 \ell})-\partial_{\tilde{\eta}}A_{0 \ell}
\end{equation}
Now, we can write the Hamiltonian containing only $A_{1l}$ and $\pi_{1 \ell}$ alone. To get to the desired form, we also make the redefinitions:
\begin{align}
    \pi_{\ell} =& \sqrt{\ell(\ell+1)\text{cosh}\tilde{\eta} }\ A_{1 \ell}\\ 
    \phi_{\ell} =& \frac{\pi_{1 \ell}}{\sqrt{\ell(\ell+1)\text{cosh}\tilde{\eta} }} \\
    [\pi_{\ell}(\tilde{\eta}),\phi_{\ell'}(\tilde{\eta}')]=&i \delta_{\ell \ell'} \delta(\tilde{\eta}-\tilde{\eta}'),
\end{align}
resulting in the Hamiltonian:
\begin{equation}
    H=\frac{1}{2}\sum_{\ell}\int d \tilde{\eta}\Big[\pi_{\ell}^2 + \text{sech} \tilde{\eta}\{\partial_{\tilde{\eta}}(\text{cosh}^{3/2} \tilde{\eta} \phi_{\ell})\}^2 + \text{cosh}^2 \tilde{\eta}\, \phi_{\ell}^2 \ell(\ell+1) \Big]
\end{equation}

\begin{figure}
    \centering
    \includegraphics[width=0.5\linewidth]{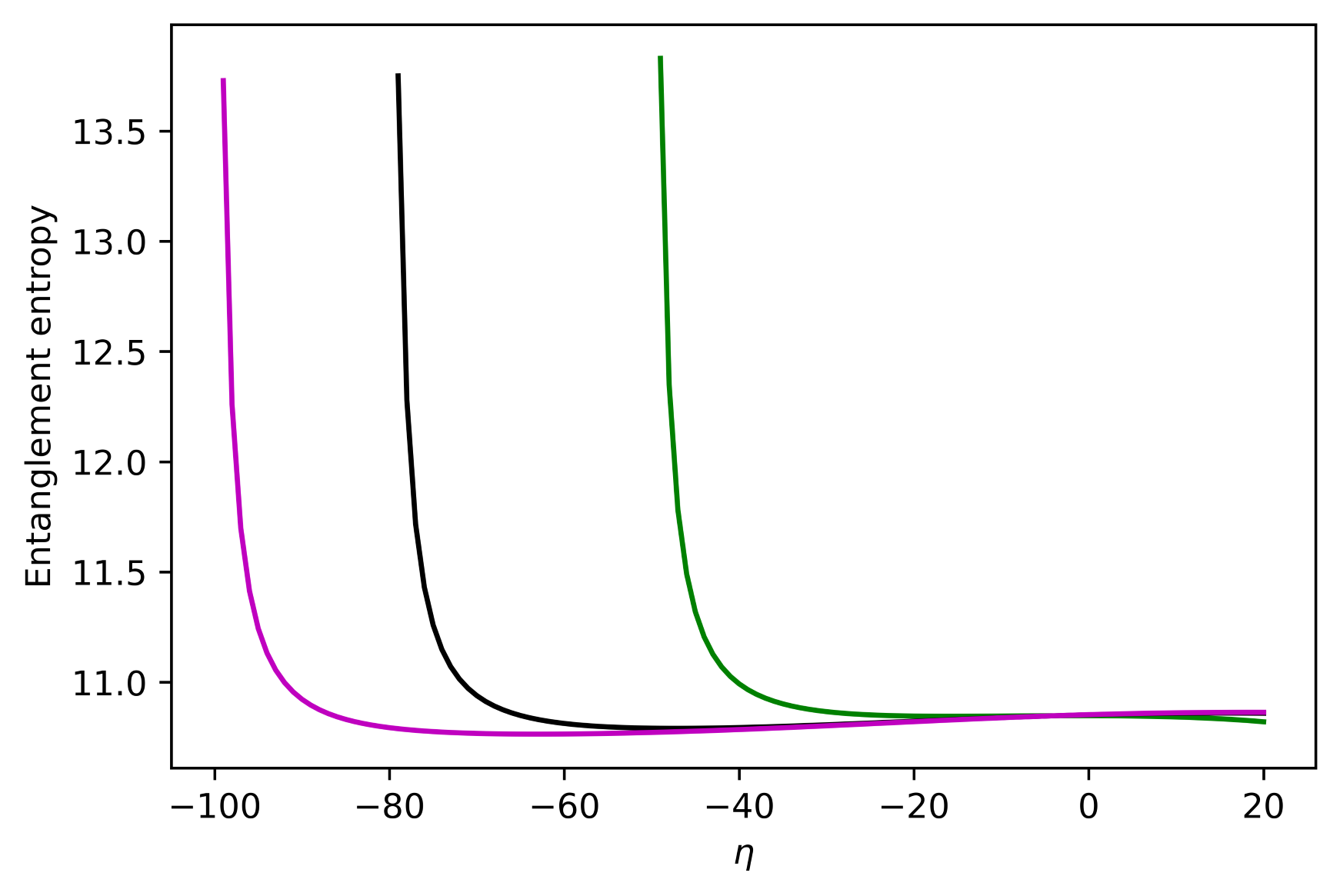}
    \caption{Entanglement entropy vs $\eta$ for gauge fields in AdS$_3$ at $L=50$ for different values of $N$. The behaviour is qualitatively same as that of scalar fields. Irrespective of how close to the boundary we start, there is an initial fall-off, after which the EE continues to be almost constant.}
    \label{gaugeL50}
\end{figure}

\begin{figure}
    \centering
    \includegraphics[width=0.5\linewidth]{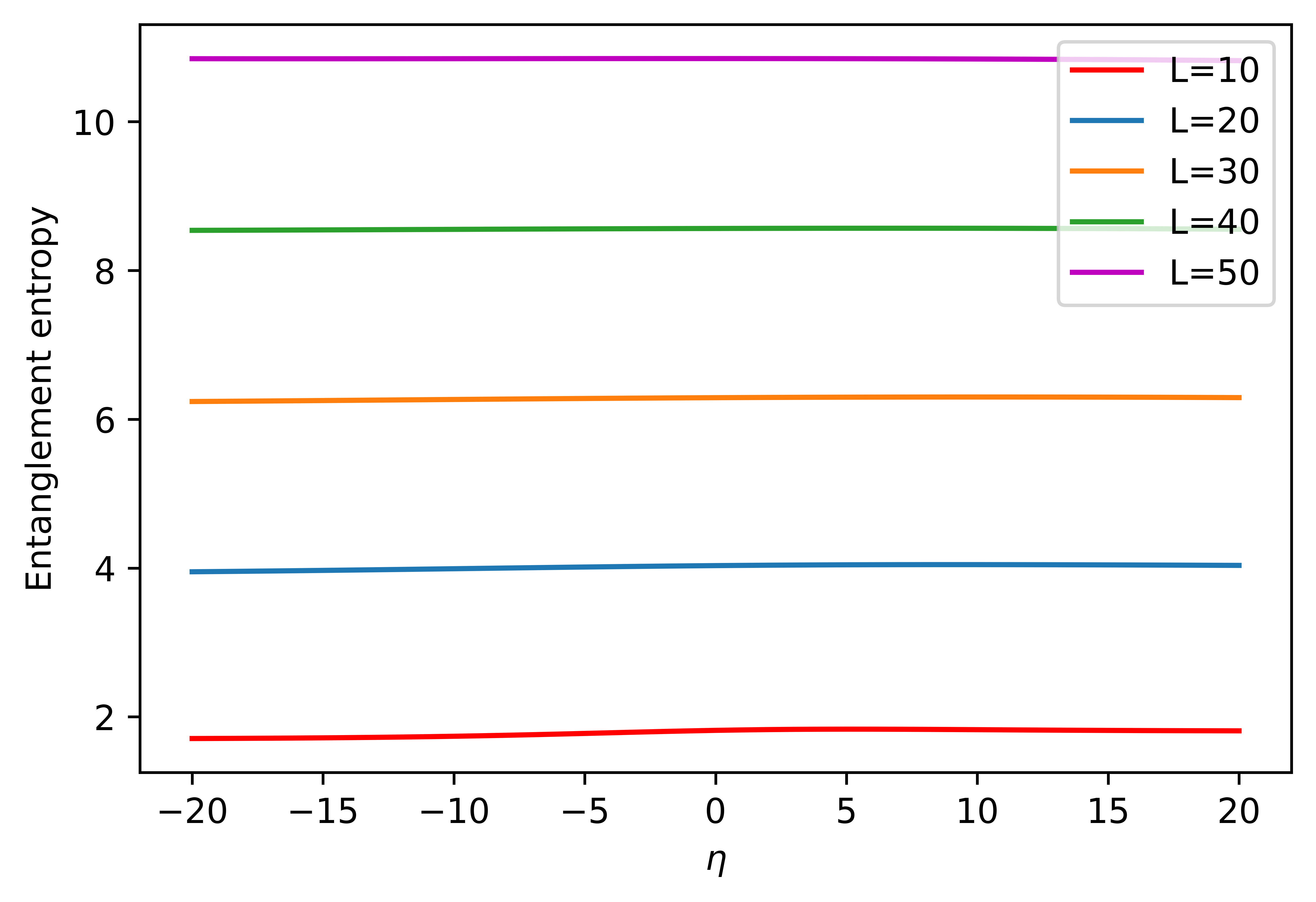}
    \caption{Entanglement entropy for gauge field inside the RT surface in global AdS3 for different $L$.}
    \label{gaugediffL}
\end{figure}

The calculation of entropy gives the entanglement entropy that has qualitatively similar behaviour to that of the scalar field case (Figures \ref{gaugeL50},\ref{gaugediffL}).

\section{Scalar fields in dS$_4$}
We start with the static patch metric for dS$_4$:
\begin{equation}
    ds^2 = -dt^2 \Big(1-\frac{r^2}{\alpha^2}\Big)+\frac{dr^2}{1-\frac{r^2}{\alpha^2}}+r^2 d\Omega^2,
\end{equation}
where $\alpha$ is the horizon. The Lagrangian density is given by \eqref{scalarfieldL}. From this, we calculate the conjugate momentum and the Hamiltonian density:
\begin{align}
    \pi =& \frac{r^2 \text{sin}\theta}{1-\frac{r^2}{\alpha^2}} \partial_0 \phi \\
    H =&\frac{1}{2} \int dr d\theta d\phi \Bigg[\pi^2\frac{1-\frac{r^2}{\alpha^2}}{r^2 \text{sin}\theta}+r^2 \text{sin}\theta\Big(1-\frac{r^2}{\alpha^2}\Big) (\partial_r \phi)^2 +\Bigg(\frac{l(l+1)}{r^2}+m^2\Bigg)\phi^2\Bigg]
\end{align}
Assuming a variable separable form for $\phi$ as before and integrating out the angular coordinates, we can simplify the Hamiltonian density.
\begin{equation}
    H= \frac{1}{2}\int dr\,  \Bigg[ \frac{1-\frac{r^2}{\alpha^2}}{r^2}\pi_{lm}^2 + r^2\Big(1-\frac{r^2}{\alpha^2}\Big)(\partial_r\phi_{lm})^2+\Big(\frac{l(l+1)}{r^2}+m^2\Big)r^2\phi_{lm}^2 \Bigg]
\end{equation}
Next, we rescale the conjugate momentum and the field:
\begin{equation}
    \tilde{\pi}_{lm}=\frac{\sqrt{1-\frac{r^2}{\alpha^2}}}{r}\pi_{lm}   \hspace{2cm} \tilde{\phi}_{lm}=\frac{r}{\sqrt{1+\frac{r^2}{\alpha^2}}}\phi_{lm}
\end{equation}

\begin{equation}
    H=\frac{1}{2}\int dr \Bigg[\tilde{\pi}_{lm}^2 +r^2\Big(1-\frac{r^2}{\alpha^2}\Big) \Bigg( \partial_r \Big[\frac{\sqrt{1-\frac{r^2}{\alpha^2}}}{r}\tilde{\phi}_{lm}\Big]\Bigg)^2 + \Big(\frac{l(l+1)}{r^2}+m^2\Big)r^2 \frac{1-\frac{r^2}{\alpha^2}}{r^2}\tilde{\phi}_{lm}^2 \Bigg]
\end{equation}
Lastly, we switch to the proper distance coordinates:
\begin{equation}
    u=\int \frac{dr}{\sqrt{1-\frac{r^2}{\alpha^2}}} = \alpha \text{sin}^{-1}\big(\frac{r}{\alpha}\big) \implies r=\alpha \, \text{sin}\Big(\frac{u}{\alpha}\Big),
\end{equation}
which holds for $r<\alpha$.
Switching to this coordinate, the Hamiltonian is:
\begin{equation}
    H=\frac{1}{2}\sum_{\ell} \int du \Bigg[ \pi_{\ell m}^2 +  L^2 \text{sin}^2\Big(\frac{u}{L}\Big) \text{cos}\Big( \frac{u}{L} \Big) \Bigg( \partial_u \Bigg\{ \frac{\text{cos}^{3/2}(\frac{u}{L})}{L\text{sin}(\frac{u}{L})}   \phi_{\ell m}\Bigg\} \Bigg)^2 + \frac{\ell(\ell+1)}{L^2 \ \text{sin}^2(\frac{u}{L})}\text{cos}^4 \Big(\frac{u}{L}\Big) \phi_{\ell m}^2 \Bigg]
\end{equation}
We see that the entropy of a spherical surface is proportional to $r^2$. Calculation is not possible for values that correspond to $r>L$. Because of the form of the metric we have chosen and the form of the proper distance variable. We see that there is a faint dependence on the value of the horizon (entanglement entropy is higher for lower values of the horizon). We also observe that there is are IR effects near the horizon which is seen by a spike in the entanglement entropy as $r$ approaches $\alpha$.

\begin{figure}
    \centering
    \includegraphics[width=0.55\linewidth]{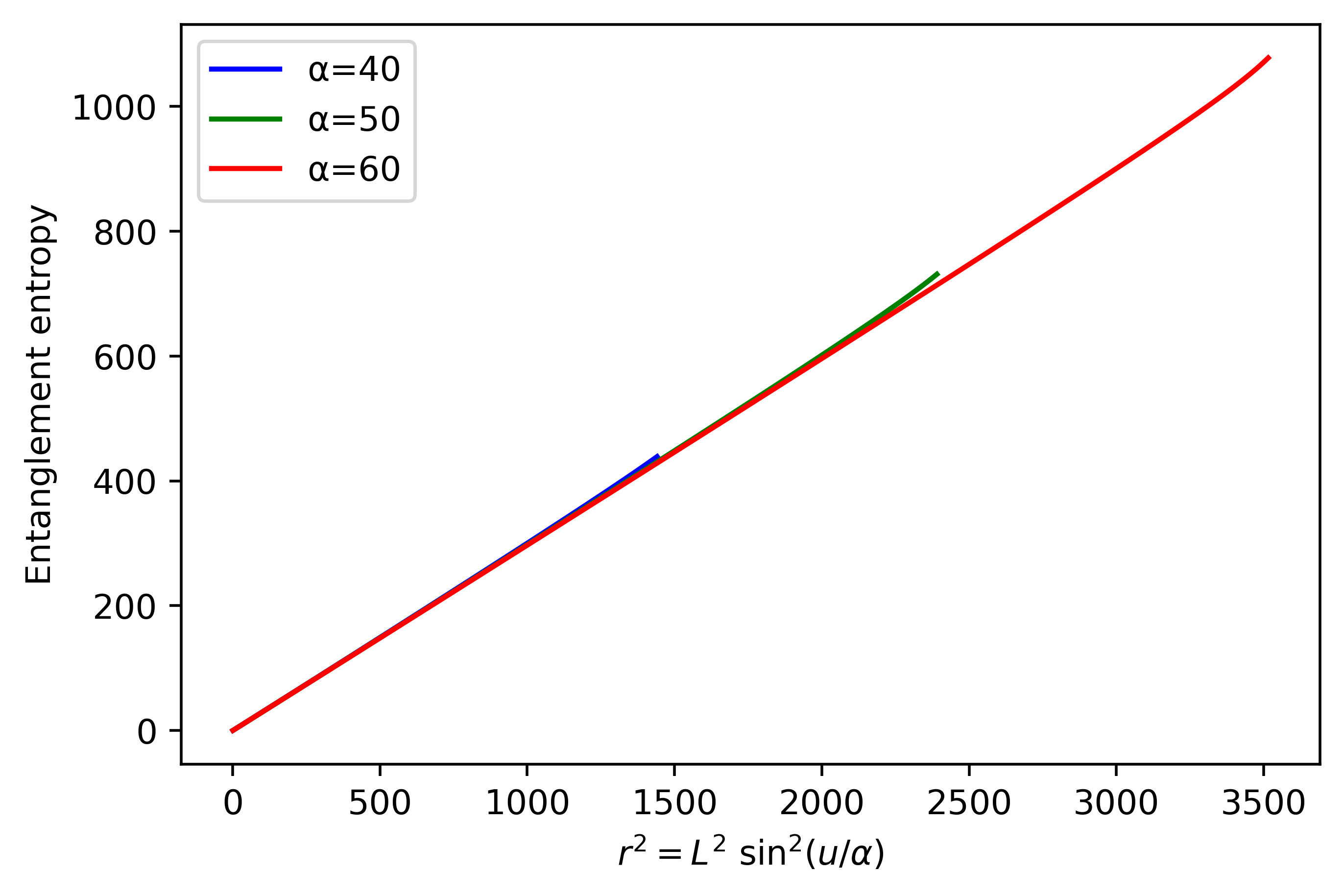}
    \caption{Entanglement entropy of scalar field dS$_4$ for different values of the horzon $\alpha$.}
\end{figure}

\section{Entanglement Negativity}
Given a bipartite state with density matrix $\rho$, the entanglement negativity is defined as \cite{Vidal:2002zz}:
\begin{equation}
\mathcal{N}(\rho)\ =\ \frac{||\rho^{T_A}||_1-1}{2},
\end{equation}
where $T_A$ corresponds to partial transpose over one of the Hilbert space and $||A||_1=\text{Tr}\sqrt{A^{\dagger}A}$ is the trace norm. Equivalently, negativity can be defined as the absolute sum of the negative eigenvalues of $\rho^{T_A}$. The logarithmic negativity is defined as:
\begin{equation}
\begin{aligned}
E_{\mathcal{N}}(\rho)\ =&\ \text{log} ||\rho^{T_A}||_1 \\
=&\ \text{log} (2\ \mathcal{N}(\rho)+1)
\end{aligned}
\end{equation}
Let $|\psi\rangle=\sum_{\alpha}c_{\alpha}|e_{\alpha}'\rangle |e_{\alpha}''\rangle$ denote the Schmidt decomposition of a pure state $|\psi\rangle$ where $c_{\alpha}$'s are the Schmidt coefficients and $e_{\alpha}^{(i)}$'s are orthonormal basis vectors. Let $\textbf{F}$ be the flip operator whose operation is defined by
\begin{equation}
\textbf{F}|e_{\alpha}'\rangle|e_{\beta}''\rangle \ =\ |e_{\beta}'\rangle |e_{\alpha}''\rangle
\end{equation}
Let $C'=\sum_{\alpha}c_{\alpha}|e_{\alpha}'\rangle \langle e_{\alpha}'|$ and $C''=\sum_{\alpha}c_{\alpha}|e_{\alpha}''\rangle \langle e_{\alpha}''|$.
Then, 
\begin{equation}
\begin{aligned}
(|\psi \rangle \langle \psi|)^{T_A}\ =&\ \sum_{\alpha \beta} c_{\alpha} c_{\beta} |e_{\alpha}'\otimes e_{\beta}''\rangle \langle e_{\beta}' \otimes e_{\alpha}''| \\
=&\ \textbf{F}\ (C' \otimes C'')
\end{aligned}
\end{equation} 
$\textbf{F}$ is a unitary operator and so, the trace norm of $(|\psi \rangle \langle \psi|)^{T_A}$ is just the trace norm of $C'\otimes C''$ which is $(\sigma_{\alpha} c_{\alpha})^2$. The negativity then can be written as \cite{Vidal:2002zz}:
\begin{equation}
\mathcal{N}(\rho)=\frac{(\sum_{\alpha} c_{\alpha})^2 - 1}{2},
\end{equation}
and the logarithmic negativity is:
\begin{equation}
E_{\mathcal{N}}(\rho)=2\, \text{log} \Big(\sum_{\alpha}c_{\alpha}\Big)
\end{equation}
$|c_{\alpha}|^2$'s are the eigenvalues of the reduced density matrix. The log negativity can be therefore written as \cite{Calabrese:2012ew}:
\begin{equation}
E_{\mathcal{N}}(\rho)=2 \, \text{log}\, \text{Tr}\rho_B^{1/2}
\end{equation}
where $\rho_B$ is the reduced density matrix. Since we already have the eigenvalues of the reduced density matrix from \eqref{evals}, we can calculate the logarithmic negativity.
\begin{equation}
\begin{aligned}
\text{Tr}\rho_B^{1/2}\ =&\ \sqrt{1-\xi} \ \sum_n \xi^{n/2}\\
=& \ \Bigg(\frac{1+\sqrt{\xi}}{1-\sqrt{\xi}}\Bigg)^{1/2}
\end{aligned}
\end{equation}

\begin{equation}
E_{\mathcal{N}}(\rho_B)\ =\ \text{log} \Bigg(\frac{1+\sqrt{\xi}}{1-\sqrt{\xi}}\Bigg)
\end{equation}

\subsection{Divergence in 3+1 d and above}
A feature of the method we are using is that the log negativity for a spherical entangling surface diverges in 3+1 dimensions. We already saw that when $l\gg N$, the eigenvalues are of the form
\begin{equation}
\xi_l(n)=\frac{g(n)}{l^2(l+1)^2}
\end{equation}
The contribution towards log negativity for large values of $\ell$ is then:
\begin{equation}\label{lneg}
\begin{aligned}
E_{\mathcal{N}(\text{large} \ell)} =& (2\ell+1) \text{log}\Bigg[\frac{\ell(\ell+1)+ \sqrt{g}}{\ell(\ell+1)-\sqrt{g}}\Bigg]\\
\approx & (2\ell+1) \frac{2\sqrt{g}}{\ell(\ell+1)}
\end{aligned}
\end{equation}
A sum over terms of the form in \eqref{lneg} does not converge. However, this problem does not show up in 2+1 dimensions since convergence is made possible by the absence of the $2 \ell+1$ factor, resulting in a fall-off of $O(\ell^{-2})$ for the log negativity as a function of $\ell$.

\subsection{$\text{E}_{\mathcal{N}}$ in 2+1 $d$}
Using the computation outlined in the previous subsection, we calculate the logarithmic negativity for scalar field in flat space and AdS$_3$ with the metric \eqref{RTmetric}. 
\begin{figure}[h]
    \centering
    \includegraphics[scale=1.2]{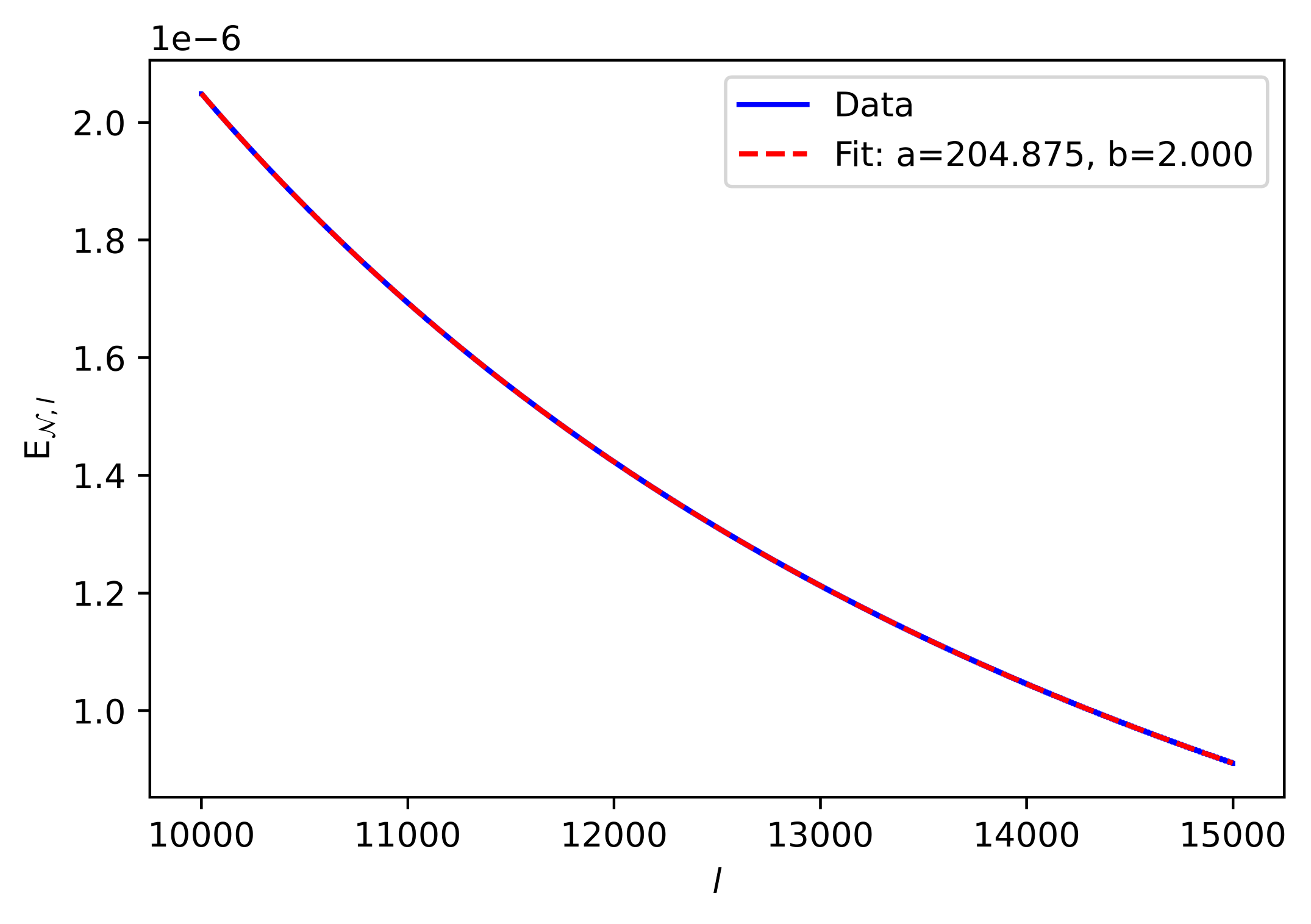}
    \caption{Fall off of logarithmic negativity for scalar fields in 2+1 d, fitted with $\frac{a}{\ell^b}$ for $N=50$ and $n=20$. We obtain a value of $b=2$, which ensures that the sum over $\ell$ converges.}
    \label{l fall off}
\end{figure}
In the calculation for flat space, figure \ref{l fall off} shows that the fall off at large values of $\ell$ is of the form $\ell^2$, ensuring convergence of the infinite sum. The log negativity increases linearly with $R$ in accordance with the area law.
\begin{figure}[h]
    \centering
    \includegraphics[scale=0.6]{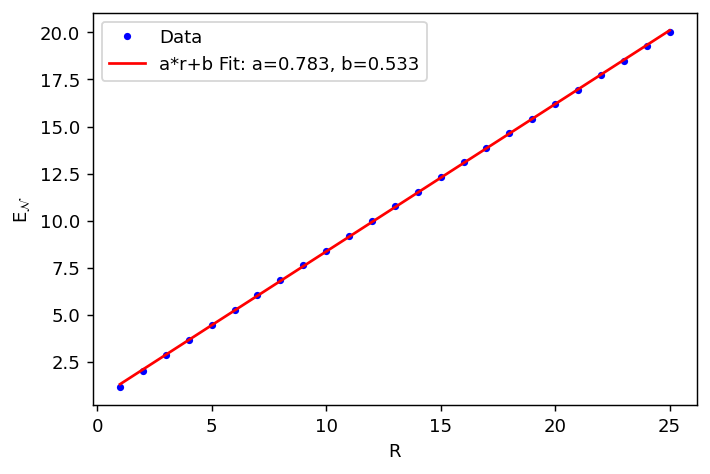}
    \caption{Log negativity in 2+1 dimensions for scalar fields in circular entangling surface is proportional to the radius $R$.}
\end{figure}

For scalar field inside RT surface in AdS$_3$, the logarithmic negativity shows similar trends as that of entanglement entropy, which is almost constant and independent of $\eta$. 
\begin{figure}
    \centering
    \includegraphics[scale=0.6]{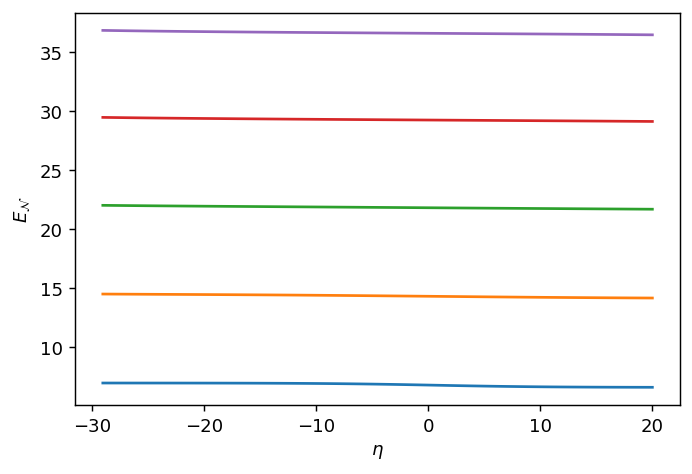}
    \caption{E$_{\mathcal{N}}$ vs $\eta$ for scalar fields inside RT surface in AdS$_3$ global coordinates.}
\end{figure}

\section{Heat Kernel Coefficients and the Universal Term}
In a $d$ dimensional relativistic QFT, for subsystem of linear size $l$, entanglement entropy has the following UV behaviour \cite{Solodukhin:2011gn,Nishioka:2018khk}
\begin{align}
    S=\left\{ \begin{array}{ll}
    a_{d-2}\left(\tfrac{l}{\epsilon}\right)^{d-2}+a_{d-4}\left(\tfrac{l}{\epsilon}\right)^{d-4}+\dots+a_1 \left(\tfrac{l}{\epsilon}\right)+(-1)^{\tfrac{d-1}{2}}c_u+\mathcal{O}(\epsilon) &\text{  for odd }d\\[3pt]
     a_{d-2}\left(\frac{l}{\epsilon}\right)^{d-2}+a_{d-4}\left(\frac{l}{\epsilon}\right)^{d-4}+\dots+(-1)^{\frac{d-2}{2}}c_u \text{log}\left(\frac{l}{\epsilon}\right)+\mathcal{O}(\epsilon^0) & \text{ for even }d
     \end{array}
    \right.
\end{align}
where $\epsilon$  is the UV cut off of the theory. The second last terms in both lines are universal and their coefficient $c_u$ contains information about conformal anomalies of the theory\cite{Casini:2004bw,Solodukhin:2008dh,Solodukhin:2011gn,Myers:2010tj}. Therefore, while other coefficients in particular do not mean anything, $S_A$ is a useful quantity to compute.
\subsection{Heal Kernel Expansion}
The heat kernel method \cite{Nishioka:2018khk} provides a systematic way to derive the entanglement entropy with all its subleading terms from the replica trick. The key to this method is calculating integrals of curvature invariants on a manifold with conical singularities\cite{Dowker:1977zj,Dowker:1994nt,Fursaev:1994in,Fursaev:1995ef,Fursaev:2013fta}. Consider the Euclidean action for free massive scalar field in spacetime $\mathcal{M}_{d-2}$.
\begin{equation}
    S=\frac{1}{2}\int d^d x \sqrt{-g}\ \Big[(\partial_{\mu} \phi)^2 +m^2 \phi^2 \Big]
\end{equation}
The entanglement entropy is calculated using the replica trick \cite{Nishioka:2018khk}.
\begin{equation}\label{replica}
    S=-\lim_{n\rightarrow 1} \partial_n (\text{log}Z_n -n\text{log}Z).
\end{equation}
Because of the replica geometry, the Laplacian factors into: $\nabla^2=\nabla_{\mathcal{C}_n}^2+\nabla_{\mathcal{M}_{d-2}}^2$ 
where $\mathcal{C}_n$ is the 2 dimensional space with a conical singularity and $\mathcal{M}_{d-2}$ is the rest of the spacetime. 
\begin{equation}
    ds_{\mathcal{C}_n}^2=dr^2+r^2 d\tau^2,
\end{equation}
$Z_n$ can be calculated by 
\begin{equation}\label{effact}
\begin{aligned}
    \text{log}Z_n=&-\frac{1}{2}\text{log\hspace{1mm}det}(-\nabla^2+m^2)\\
    =&-\frac{1}{2}\text{tr\hspace{1mm}log} (-\nabla^2+m^2)\\
    =&\frac{1}{2}\int_{\epsilon^2}^\infty \frac{ds}{s} \text{tr}[e^{-s(-\nabla^2+m^2)}-e^{-s}],
\end{aligned}
\end{equation}
where $s$ is the Schwinger parameter. 
where $0\leq r$ and $0\leq \tau\leq 2\pi n$. Using the eigenfunctions of $\nabla_{\mathcal{M}_{d-2}}^2$ and $\nabla_{\mathcal{C}_n}^2$, entanglement entropy is calculated\cite{Nishioka:2018khk}.
\begin{equation}
    S=\frac{\pi \text{Vol}(\mathcal{M}_{d-2})}{3} \int_{\epsilon^2}^\infty ds \frac{e^{-sm^2}}{(4 \pi s)^{d/2}}
\end{equation}
Given that equation \eqref{effact} is of the form
\begin{equation}
    \text{log} Z_n=\frac{1}{2}\int_{\epsilon^2}^\infty \frac{ds}{s} \text{tr}K_{\mathcal{M}_n}(s)e^{-sm^2},
\end{equation}
the heat kernel operator $K_{\mathcal{M}_n}(s)$ has the expansion around $s=0$ of the form
\begin{equation}
    \text{tr} K_{\mathcal{M}_n}(s)=\frac{1}{(4 \pi s)^{d/2}} \sum_{i=0}^\infty a_i (\mathcal{M}_n) s^i,
\end{equation}
where $a_i$'s are the heat kernel coefficients, which can be calculated using quantities related to the curvature of the hypersurface we are interested in. In the $n\rightarrow 1$ limit, the heat kernel coefficients have the expansion:
\begin{align}
    a_i\ =&\ a_i^{\text{bulk}}+(1-n)a_i^{\Sigma}+O((1-n)^2)\\
    a_i^{bulk}(\mathcal{M}_n)\ =&\  n\, a_i^{bulk}(\mathcal{M}_1)
\end{align}
The $a_i^{\Sigma}$ from the contributions due to the conical singularity. Due to \eqref{replica}, the equation for the entanglement entropy will contain only $a_i^{\Sigma}$'s.
\begin{equation}\label{hkentropy}
    S=\frac{1}{2(4 \pi)^{d/2}} \sum_{i=0}^{\infty} \frac{a_i^{\Sigma}}{m^{2i-d}}\Gamma \Big(i-\frac{d}{2},m^2 \epsilon^2 \Big),
\end{equation}
where $\Gamma(t,x)=\int_x^{\infty} ds\, s^{t-1}e^{-s}$ is the incomplete Gamma function. The expressions for the first few heat kernel coefficients are:
\begin{align}
    a_0^{\Sigma}\ =&\ 0\\
    a_1^{\Sigma}\ =&\ \frac{2 \pi}{3} \int_\Sigma 1\\
    a_2^{\Sigma}\ =&\ \frac{\pi}{9} \int_{\Sigma} \Bigg[\frac{1}{5}\Big(2\mathcal{R}_{abab}-\mathcal{R}_{aa}-2\mathcal{K}^a_{\mu \nu}\mathcal{K}^{a\, \mu \nu}+\frac{1}{2}(\mathcal{K}^{a\, \mu}_{\mu})^2\Big)+\mathcal{R}\Bigg],
\end{align}
where 
\begin{align}
    \mathcal{R}_{abab}&=R_{\mu \nu \rho \sigma}n_{a}^{\mu}n_{b}^{\nu} n_{a}^{\rho}n_{b}^{\sigma}\\   
    \mathcal{R}_{aa}&=R_{\mu \nu} n_a^{\mu} n_a^{\nu}.
\end{align} 
$\mathcal{K}^a_{\mu \nu}$ is the extrinsic curvature associated with the normal vector $n_a^\mu$ on $\Sigma$. 
\begin{equation}
    K^n_{\mu \nu}=h^{\rho}_{\mu} h^{\sigma}_{\nu} \nabla_{\rho}n_{\sigma},
\end{equation}
where $h$ is the induced metric:
\begin{equation}
    h_{\nu}^{\mu}=\delta_{\nu}^{\mu}+n^{\mu}n_{\nu}
\end{equation}
From \eqref{hkentropy}, we see that $a_1^{\Sigma}$ shows up in the term proportional to $\frac{R^{d-2}}{\epsilon^{d-2}}$ and $a_2^{\Sigma}$ shows up in the next subleading term which is of the form $\text{log}(\frac{R}{\epsilon})$. The above formula for the second heat kernel coefficient can be used to calculate the coefficient of the universal term in the UV expansion of entanglement entropy for the various geometries we have considered.
It is also possible to compute the effect of mass on the entanglement entropy. To leading order, for \eqref{hkentropy}, we can calculate
\begin{equation}\label{masskernel}
\begin{aligned}
    S_A=&\frac{1}{48 \pi} A m^{2} \int_{m^2\epsilon^2}^{\infty} ds \frac{1-s^2}{s^2}\\
    =& \frac{1}{48 \pi} A \Bigg[\frac{1}{\epsilon^2}-m^2\Big(\int_{m^2\epsilon^2}^1\frac{1}{s} ds+\int_1^{\infty}\frac{1}{s} ds\Big)\Bigg]\\
    =& \frac{A}{48 \pi}\Big[\frac{1}{\epsilon^2} +2m^2 \text{log}(m\epsilon)+m^2 \gamma\Big] 
\end{aligned}
\end{equation}
where $\gamma$ is the Euler-Mascheroni constant defined by
\begin{equation}
    \gamma = \lim_{n\rightarrow\infty}\Big(-\text{log}n +\sum_{k=1}^n \frac{1}{k}\Big)
\end{equation}
The expansion for higher order terms in equation \eqref{masskernel} is divergent. The exact expression for the mass-dependent terms is found using dimensional regularisation \cite{Nishioka:2018khk}.
\subsubsection{Spherical hypersurface in 3+1 d flat space}
For a 2-sphere in four spacetime dimensions, there are two normals, one in the timelike direction and one in the radial direction. For the timelike normal vector, the extrinsic curvature vanishes. We call the radial normal vector $n^{\mu}=(0,1,0,0)$. The non vanishing components of the extrinsic curvature associated with $n^{\mu}$ are:
\begin{equation}
    \mathcal{K}^n_{\theta \theta}=R ; \ \ \ \mathcal{K}^n_{\phi \phi}=R\, \text{sin}^2 \theta
\end{equation} 
This gives the value of the third heat kernel coefficient to be:
\begin{equation}
\begin{aligned}
    a_2^{\Sigma}\ =&\ -\frac{\pi}{45}\frac{2}{R^2}4 \pi R^2 \\
    =& \ -\frac{8 \pi^2}{45}
\end{aligned}
\end{equation}
Then, the second term in the expansion of \eqref{hkentropy} gives:
\begin{equation}
\begin{aligned}
    a_2^{\Sigma} \int_{\epsilon^2}^{\infty} ds\, m^2 \frac{1}{2 (4 \pi)^2} \frac{1}{m^2\, s} \ =&\ \text{Log divergent term}+\frac{8 \pi^2}{45}\frac{1}{2(4\pi)^2}2\text{log}\epsilon\\
    =&\ -\frac{1}{90}\text{log}\frac{1}{\epsilon}
\end{aligned}   
\end{equation}
This gives a value of $-\frac{1}{90}$ for the coefficient of the log term which was also obtained numerically in \cite{Lohmayer:2009sq}.
\subsubsection{Spherical hypersurface in global AdS$_4$}
In this case too, there are two normals, one in the timelike direction and one in the radial direction. Since the spacetime is static, once again, the extrinsic curvature vanishes  but the curvature terms are now non zero. We write the vectors as $n_1^{\mu}=\Big(-\tfrac{1}{\sqrt{1+\tfrac{r^2}{L^2}}},0,0,0\Big)$ and $n_2^{\mu}=(0,\frac{1}{\sqrt{1+\frac{r^2}{L^2}}},0,0)$. The non vanishing components of the extrinsic curvature are:
\begin{equation}
    K^{n_2}_{\theta \theta}=R\sqrt{1+\frac{R^2}{L^2}}; \ K^{n_2}_{\phi \phi}=R\, \sin^2\theta\sqrt{1+\frac{R^2}{L^2}}
\end{equation}
Moreover, the $\mathcal{R}_{aa}$ term here is non-zero.
\begin{equation}
\begin{aligned}
    \mathcal{R}_{aa}=&\mathcal{R}_{rr}n_2^r n_2^r+\mathcal{R}_{tt}n_1^t n_1^t\\
    =&-\frac{3}{L^2}+\frac{3}{L^2}\\
    =& 0
\end{aligned}
\end{equation}

\begin{equation}
\begin{aligned}
    \mathcal{R}_{abab}=&\mathcal{R}_{rtrt}(n_2^r)^2(n_1^t)^2+\mathcal{R}_{trtr}(n_1^r)^2(n_2^t)^2 \\
    =\frac{2}{L^2}
\end{aligned}
\end{equation}

Now substituting these into the equation for $a_2^{\Sigma}$:
\begin{equation}
\begin{aligned}
    a_2^{\Sigma}=&\frac{\pi}{9}4 \pi R^2 \Bigg[\frac{1}{5} \Big(\frac{4}{L^2}-\frac{2(1+\tfrac{R^2}{L^2})}{R^2}\Big)-\frac{12}{L^2}\Bigg]\\
    =& \frac{4\pi R^2}{9}\Big[\frac{-58}{L^2}-\frac{2}{5 R^2}\Big]
\end{aligned}
\end{equation}
In the $L\rightarrow \infty$ limit, this coefficient reduces the flat space value of $-1/90$.
\subsubsection{Constant $\eta$ surfaces in AdS$_3$}
In this case, the universal piece is a term independent of $\epsilon$. However, the method of using the heat kernel coefficient cannot be used to obtain the term independent of $\epsilon$. The first heat kernel coefficient can still be used to derive the term with the leading divergence:
\begin{equation}
\begin{aligned}
    S=& \frac{1}{2(4\pi)^{3/2}}a_1^{\Sigma}m\int_{m^2\epsilon^2}^{\infty} ds \, s^{-3/2}e^{-s}.
\end{aligned}
\end{equation}
From this, we can find
\begin{equation}
    S_{\mathcal{O}(\epsilon^{-1})}= \frac{1}{12\sqrt{\pi}}\frac{L}{\epsilon} V\ ,
\end{equation}
where $V$ is the factor from integrating over the constant $\eta$ slice which is infinite unless there is a cut-off on the $x$ coordinate which otherwise runs from $-\infty$ to $\infty$. The result is directly proportional to $L$, which indeed agrees with what we have obtained from the numerical calculations.
\section{Conclusion}
We have performed numerical calculations of the entanglement entropy and logarithmic negativity for fields in differnt backgrounds, namely AdS$_4$ and AdS$_3$. We see that the area-law behavior is most easily seen when the proper distance coordinate is discretized and used in the calculation. We have extended the use of vector spherical harmonics to global AdS$_4$ to calculate the entanglement entropy for gauge fields. Moreover, we have found the appropriate coordinates that will allow us to carry out these calculations for fields inside an RT surface in AdS$_3$. While the calculation is more or less straightforward for scalar fields once we have the coordinates, we see that we need to make a very non-trivial gauge choice in order to integrate out the $x$ coordinate out of the Hamiltonian for gauge fields. We have also calculated the coefficient of the universal term in the entanglement entropy for scalar field in global AdS$_4$ using the heat kernel method. This coefficient contains an term that depends on $L$ such that the coefficient reduces to the correct flat space value in the $L\rightarrow\infty$ limit.

\section*{Acknowledgements}
We would like to thank Dawood Kothawala for helpful discussions. We would also like thank Shanmugapriya P. for collaboration in the initial stages of the project and useful discussions. SP acknowledges the support of Center for Scientific and Industrial Research (CSIR) through the Senior Research Fellowship.

\appendix
\section{AdS-Rindler Coordinates}\label{appA}
For a $d+1$ dimensional AdS spacetime embedded in $d+2$ dimensional space $\mathbb{R}^{2,d}$ with coordinates $T_1,T_, \, X_1, \, \cdots X_{d-1}$:
\begin{equation}
    -T_1^2-T_2^2+\sum_{i=1}^{d-1}X_i^2=-L^2\ ,
\end{equation}
AdS-Rindler coordinates are defined as follows:
\begin{align}
    T_1 \ =\ & \sqrt{\rho^2-L^2} \ \text{cosh}\Big(\frac{\tau}{L}\Big)\text{sinh}\Big(\frac{\eta}{L}\Big)+\rho \ \text{cosh} \Big(\frac{x}{L}\Big) \text{cosh}\Big(\frac{\eta}{L}\Big)\\    
    T_2 \ =\ &\sqrt{\rho^2-L^2} \ \text{sinh}\Big(\frac{\tau}{L}\Big)\\
    X_d \ = \ & \sqrt{\rho^2-L^2} \ \text{cosh} \Big(\frac{\tau}{L}\Big) \text{cosh}\Big(\frac{\eta}{L}\Big)+\rho \ \text{cosh}\Big(\frac{x}{L}\Big) \text{sinh}\Big(\frac{\eta}{L}\Big)\\
    X_1 \ = \ & \rho \sinh\Big(\frac{x}{L}\Big)\cos \theta_1\\
    X_2 \ = \ & \rho \sinh\Big(\frac{x}{L}\Big)\sin \theta_1 \cos \theta_2 \\  
    \vdots \notag \\
    X_{d-1} \ = \ & \rho \, \sinh \Big(\frac{x}{L}\Big) \sin \theta_1 \sin\theta_2 \cdots \sin \theta_{d-1}
\end{align}
The metric can be found to be:
\begin{equation}
    ds^2=-d \tau^2(\rho^2-L^2)+\frac{d\rho^2}{\rho^2-L^2}+\rho^2 (dx^2+\sinh^2x \, d\Omega_{d-1}^2)
\end{equation}
$\rho=L$ surfaces are geodesics. These coordinates are therefore a natural choice to work with RT surfaces. It is to be noticed how $\eta$ is not present at all in the metric, but is rather just a parameter in the definitions of the Rindler coordinates. Rindler coordinates do not cover the whole of AdS and the value of $\eta$ determines the size of the Rindler patch. The definitions of the coordinates can be used to find expressions for global coordinates in terms of Rindler coordinates.

\begin{align}\label{rindlerbegin}
    r \ = \ &\sqrt{\rho^2\text{sinh}^2\Big(\frac{x}{L}\Big)+\Big(\sqrt{\rho^2-L^2} \ \text{cosh}\Big(\frac{\eta}{L}\Big) \text{cosh} \Big(\frac{\tau}{L}\Big)+\rho \text{cosh}\Big(\frac{x}{L}\Big) \text{sinh}\Big(\frac{\eta}{L}\Big)\Big)^2}\\
    t \ = \ & L \ \text{tan}^{-1} \bigg(\frac{\text{sinh}(\tau/L) \sqrt{\rho^2-L^2}}{\rho \text{cosh}(x/L) \text{cosh} (\eta/L)+\sqrt{\rho^2-L^2} \text{cosh}(\tau/L)\text{sinh}(\eta/L)}\bigg)\\
    \phi \ = \ & \ \text{tan}^{-1} \bigg( \frac{\rho \ \text{sinh}(x/L)}{\sqrt{\rho^2-L^2} \text{cosh}(\tau/L) \text{cosh}(\eta/L) + \rho \text{cosh}(x/L)\text{sinh}(\eta/L)} \bigg) \label{rindlerend}
\end{align}
$\eta=0$ gives rise to the most familiar form of Rindler coordinates, which covers half of the global AdS patch.

\section{RT surface in AdS$_3$}
We look at the RT surface in AdS$_3$ using global coordinates. The metric is given by equation \eqref{globmetric}. On a constant time slice,
the length of the metric will be given by
\begin{equation}
    l=\int \sqrt{\frac{r'^2}{1+\frac{r^2}{L^2}}+r^2}\, d\phi,
\end{equation}
where $r'$ denotes derivative of $r$ with respect to $\phi$. We can extract a conserved quantity from the integrand of $l$ as follows.
\begin{equation}
\begin{aligned}
     r_*=&\frac{\frac{r'^2}{1+\frac{r^2}{L^2}}}{\sqrt{\frac{r'^2}{1+\frac{r^2}{L^2}}+r^2}}- \sqrt{\frac{r'^2}{1+\frac{r^2}{L^2}}+r^2} \\
     =& -\frac{r^2}{\sqrt{\frac{r'^2}{1+\frac{r^2}{L^2}}+r^2}}
\end{aligned}
\end{equation}

\begin{equation}
    r_*^2\Big(\frac{r'^2}{1+\frac{r^2}{L^2}}+r^2\Big)=r^4
\end{equation}
From this, we obtain an expression for $r'$:
\begin{equation}
    r'=\Big[\Big(\frac{r^4}{r_*^2}-r^2\Big)\Big(1+\frac{r^2}{L^2}\Big)\Big]^{1/2}
\end{equation}
Thus, we see that $r_*$ is the turning point of the RT surface or the point deepest into the bulk. If $\theta$ is the angular size of the boundary interval, then
\begin{equation}\label{angsize}
\begin{aligned}
    \frac{\theta}{2}=&\int_{r_*}^{r_{\infty}} \frac{dr}{\sqrt{\Big(\frac{r^4}{r_*^2}-r^2\Big)\Big(1+\frac{r^2}{L^2}\Big)}}\\
    =& \text{tan}\Big(\frac{1}{r_*}\Big) \text{ as } r_{\infty} \rightarrow \infty
\end{aligned}
\end{equation}
Now, let us look at RT surfaces from the perspective of the AdS-Rindler coordinates. Substituting the value of $r_*$ into the LHS of \eqref{rindlerbegin} and setting $\rho=L$, we have
\begin{equation}
\begin{aligned}
    L^2 \text{cot}^2\Big(\frac{\theta}{2}\Big)=&L^2 \text{sinh}^2 \Big(\frac{x}{L}\Big)+L^2\text{cosh}^2\Big(\frac{x}{L}\Big)\Big(\text{cosh}^2 \Big(\frac{\eta}{L}\Big)-1\Big) \\
    =& -L^2+L^2\text{cosh}^2\Big(\frac{x}{L}\Big)\text{cosh}^2\Big(\frac{\eta}{L}\Big)
\end{aligned}
\end{equation}
From the definition of global and AdS-Rinddler coordinates in terms of the embedding coordinates, we alo have
\begin{equation}
    x=L\text{cosh}^{-1} \Bigg(\frac{\sqrt{r^2+L^2}}{(r^2\text{cos}^2\phi+L^2)^{1/2}}\Bigg)
\end{equation}
In writing down the LHS of equation \eqref{angsize}, we had taken $\phi=0$ to be the midpoint of the boundary interval without loss of generality. Plugging in $\phi=0$ in the above equation gives $x=0$. This gives us
\begin{equation}
    \text{cosec}\Big(\frac{\theta}{2}\Big)=\text{cosh}\Big(\frac{\eta}{L}\Big)
\end{equation}

\section{Entanglement Negativity of Two Coupled Harmonic Oscillators}
Here, we rigorously derive the entanglement negativity for a system of two coupled harmonic oscillators.
We recall that making a change of variables, we were able to rewrite the Hamiltonian of two coupled harmonic oscillators as that for two uncoupled oscillators with frequencies $k_0$ and $k_0+2k_1$.

\begin{equation}
\omega_+^2=k_0 \hspace{2cm }\omega_-^2 = k_0 +2k_1.
\end{equation}

Thus the system decomposes into two independent oscillators.

For an oscillator of frequency $\nu$:
\begin{equation}
a(\nu)
=
\frac{1}{\sqrt{2 \nu}}
\left(
\nu q + i p
\right).
\end{equation}

Hence,
\begin{align}
a_1 &= \frac{1}{\sqrt{2 \omega}}(\omega x_1 + i p_1),\\
a_2 &= \frac{1}{\sqrt{2 \omega}}(\omega x_2 + i p_2),
\end{align}

and for the relative mode:
\begin{equation}
a_-
=
\frac{1}{\sqrt{2 \omega_-}}
\left(
\omega_- x_- + i p_-
\right).
\end{equation}

Substituting $x_-=\frac{x_1-x_2}{\sqrt{2}}$ and
$p_-=\frac{p_1-p_2}{\sqrt{2}}$ and simplifying gives:

\begin{equation}
a_-
=
\frac{1}{\sqrt{2}}
\left[
\left(
\frac{\sqrt{\omega_-}}{2\sqrt{\omega_+}}
+
\frac{\sqrt{\omega_+}}{2\sqrt{\omega_-}}
\right)(a_1-a_2)
+
\left(
\frac{\sqrt{\omega_-}}{2\sqrt{\omega_+}}
-
\frac{\sqrt{\omega_+}}{2\sqrt{\omega_-}}
\right)(a_1^\dagger-a_2^\dagger)
\right].
\end{equation}

Define the squeezing parameter $r$:
\begin{align}
\cosh r &= \frac12\left(\sqrt{\frac{\omega_-}{\omega_+}}+\sqrt{\frac{\omega_+}{\omega_-}}\right),\\
\sinh r &= \frac12\left(\sqrt{\frac{\omega_-}{\omega_+}}-\sqrt{\frac{\omega_+}{\omega_-}}\right),
\end{align}
so that
\begin{equation}
e^{2r}=\frac{\omega_-}{\omega_+}.
\end{equation}

Then:
\begin{equation}
\boxed{
a_Y
=
\cosh r\,\frac{a_1-a_2}{\sqrt{2}}
+
\sinh r\,\frac{a_1^\dagger-a_2^\dagger}{\sqrt{2}}
}
\end{equation}

The true vacuum satisfies:
\begin{equation}
a_- |{\Psi_0}\rangle=0 .
\end{equation}

The normalized solution is the two-mode squeezed vacuum:
\begin{equation}
|{\Psi_0}\rangle
=
\frac{1}{\cosh r}
\exp\!\left(
\tanh r\, a_1^\dagger a_2^\dagger
\right)
|{0}\rangle_1|{0}\rangle_2
\end{equation}

Expanding:
\begin{equation}
|\Psi_0\rangle
=
\sqrt{1-\lambda^2}
\sum_{n=0}^{\infty}
\lambda^n
|n,|n\rangle ,
\qquad
\lambda=\tanh r ,
\end{equation}
where $|n,m\rangle :=|n\rangle_1\,|m\rangle_2$.
The density matrix is
\begin{equation}
\rho = |{\Psi_0}\rangle \langle {\Psi_0}|.
\end{equation}
Thus
\begin{equation}
\rho
=
(1-\lambda^2)
\sum_{n,m=0}^{\infty}
\lambda^{n+m}
|{n,n}\rangle \langle {m,m}|
\end{equation}
This is a pure but entangled state.
We define the partial transpose with respect to subsystem $A$ (oscillator 1):
\begin{equation}
\left(
|{i}\rangle_A|{j}\rangle_B
\langle{k}|_A \langle{l}|_B
\right)^{T_A}
=
|{k}\rangle_A |{j}\rangle_B
\langle{i}|_A\langle{l}|_B .
\end{equation}
Applying this to $\rho$, we obtain

%
%
\begin{equation}
\rho^{T_A}
=
(1-\lambda^2)
\sum_{n,m=0}^{\infty}
\lambda^{n+m}
|{m,n}\rangle \langle{n,m}|
\end{equation}
Thus
\begin{equation}\label{ptranspose}
    \rho^{T_A}\,|n,m \rangle=\lambda_n \lambda_m  |m,n \rangle,
\end{equation}
where $\lambda_n=\sqrt{1-\lambda^2}\ \lambda^{n}$. Comparing with equation \eqref{evals}, we see that $\lambda^2$ is the same quantity that we defined in section \ref{srednicki} as $\xi$.  Thus, we have
\begin{equation}
    \lambda_n=\sqrt{1-\xi}\ \xi^{\frac{n}{2}}\ .
\end{equation}
Equation \eqref{ptranspose} shows that $(|n,m\rangle-|m,n\rangle)$ will be eigenfunctions of $\rho^{T_A}$ with eigenvalues $-\lambda_n \lambda_m$. These are the negative eigenvalues of $\rho^{T_A}$. Summing over the absolute values of these gives the negativity.
\begin{align*}
    \mathcal{N}=&\frac{1}{2}\sum_{n\neq m}\lambda_n \lambda_m 
    =\frac{1}{2}\Bigg[\Big(\sum_{n}\lambda_n\Big)^2 -\sum_{n}\lambda_n^2\Bigg]
=\frac{\sqrt{\xi}}{1-\sqrt{\xi}}
\end{align*}
From this we find the log negativity to be
\begin{equation}
\begin{aligned}
    E_{\mathcal{N}}=\log(2 \mathcal{N}+1)
    =\log \Bigg(\frac{1+\sqrt{\xi}}{1-\sqrt{\xi}}\Bigg)\ .
\end{aligned}
\end{equation}

\nocite{*}
\bibliographystyle{JHEP.bst}
\bibliography{biblio}





\end{document}